\documentclass[preprintnumbers,amsmath,amssymb]{revtex4}
\usepackage{graphicx}
\usepackage{dcolumn}
\usepackage{bm}

\begin{document}

\title{
Topological approach for laser acceleration investigation
}

\author{Toshimasa~Morita}

\affiliation{Quantum Beam Science Research Directorate,
National Institutes for Quantum and Radiological Science and Technology,
8-1-7 Umemidai, Kizugawa, Kyoto 619-0215, Japan}


\begin{abstract}
An approach using topology reveals a new understanding and knowledge of
laser particle acceleration.
Laser pulse irradiation on a thin-foil target is examined using
two-dimensional particle-in-cell simulations.
Through topology, laser ion acceleration schemes are classified depending on
whether the deformed target shape after laser irradiation is homeomorphic with
respect to the initial target.
High energy ions are generated with high-efficiency when the deformed target
shape is non-homeomorphic.
In this case, the target thickness greatly affects the obtained ion energy.
However, when the deformed target shape is homeomorphic, the obtained ion
energy is low, and the effect of the target thickness is small.
\end{abstract}

\maketitle

\section{Introduction}

Recently, significant progress has been achieved in compact laser systems,
and laser ion acceleration is a compelling application of
high-power compact lasers \cite{BWE,DNP}.
If a compact laser system is able to generate ions with sufficiently
high energy and quality, low-cost compact accelerators would become feasible.
However,
the ion energies achieved thus far are insufficient for some applications,
such as hadron therapy \cite{CLK,SNV}.
In addition,
such applications require the ion beam to have a narrow energy spread,
high quality, and a sufficient number of ions \cite{ROT,ESI,BEE}.
Therefore, it is important to study the best conditions for generating ions
with higher energies and of higher quality
\cite{BWP,DL,HSM,PPM,PRK,SPJ,Toncian,TM1,TM2}.
Moreover, in laser ion accelerator technologies,
the required energy and number of ions vary depending on the application, i.e.,
it is necessary to establish a method to generate specified particular amount
of energy and number of ions through laser ion acceleration.
Therefore, what is important for laser acceleration investigation from now on
is research on methods to control the generated ion energy and number.
To this end, we need to understand the laser particle acceleration phenomenon
in sufficient detail.
In this paper, we have added a new understanding to laser acceleration
using the fundaments of topology.

Topology is used in almost all the fields of mathematics.
It is one of the fundamental theories that support not only the field of
geometry but also general mathematics.
Additionally, it is one of the fundamental theories of physics and engineering,
which are based on mathematics.
In this paper, we study the laser particle acceleration phenomenon from
the viewpoint of topology.

In Sec. \ref{topo},
how to treat the laser acceleration phenomena with topology is shown.
In Secs. \ref{h-para} and \ref{h-resu},
we present some considerations of laser particle acceleration using the
hydrogen foil target.
In Sec. \ref{ch2},
the polyethylene foil target case is shown.
In the final section, the main results of our study are summarized.

\section{Topology and laser acceleration} \label{topo}

\begin{figure}[tbp]
\includegraphics[clip,width=9.0cm,bb=8 5 462 255]{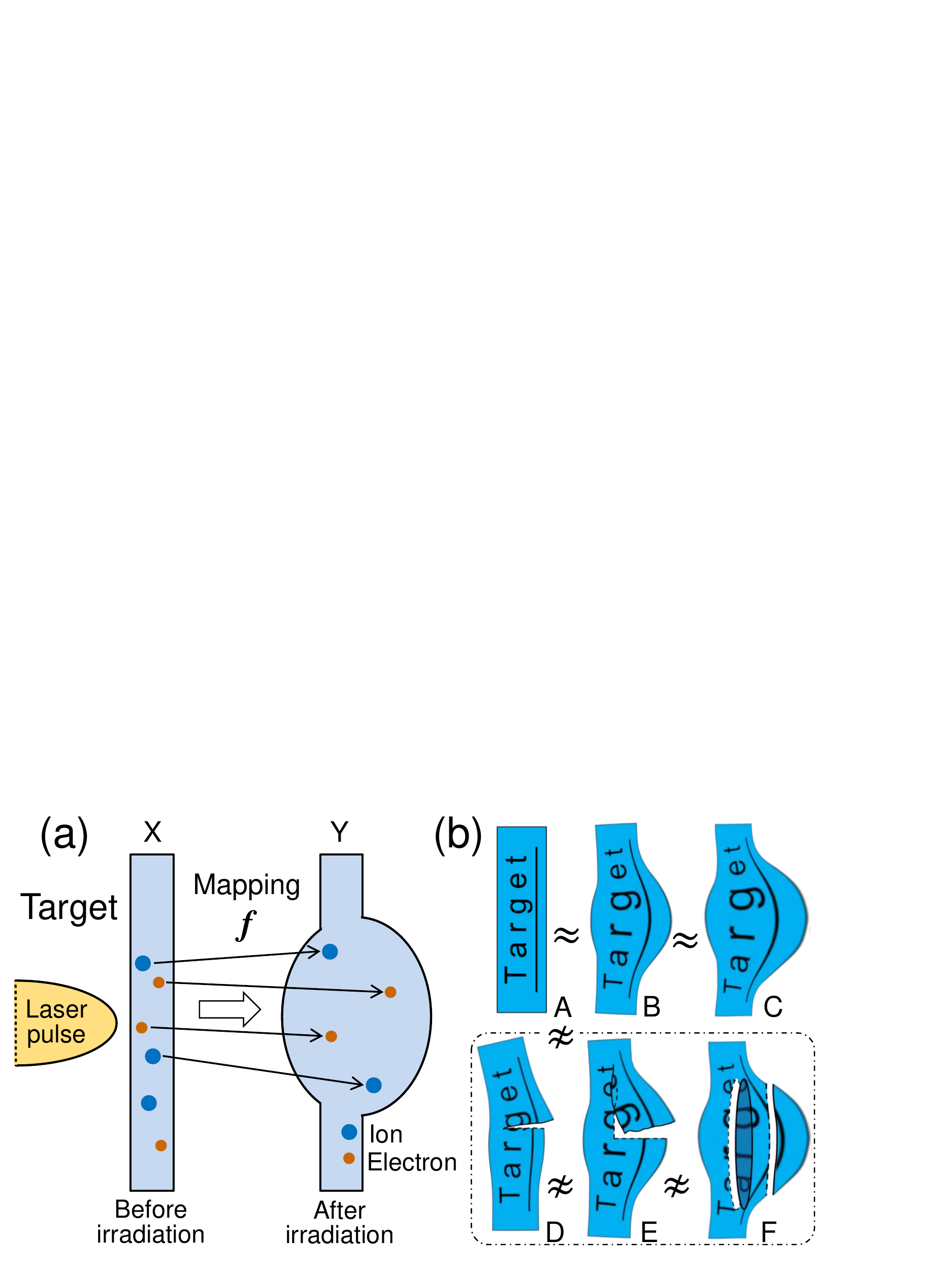}
\caption{
(a) The laser particle acceleration process is understood
as a bijective mapping $f:X \rightarrow Y$.
(b) $A$, $B$, and $C$ are homeomorphic ($A\approx B\approx C$).
$D$, $E$, and $F$ are non-homeomorphic to $A$ and to each other
($A \not\approx D \not\approx E \not\approx F$).
}
\label{fig:fig-mp}
\end{figure}

First, we consider the target before laser irradiation, and the ``target'' after
laser irradiation which is precisely composed a cloud of ions and electrons.
This can be regarded as a mapping $f$ from the target before laser irradiation,
$X$, to that after laser irradiation, $Y$, (Fig.\ref{fig:fig-mp}(a)).
\begin{equation}
f:X \rightarrow Y.
\label{map}
\end{equation}
It is assumed that no nuclear reaction occurs; therefore,
an ion and electron do not disappear or split into two or generate.
Therefore,
\begin{enumerate}
\item
The ions and electrons in the target before laser irradiation are always
present somewhere in the target after irradiation; i.e.,
one-to-one mapping $=$ injection,
\begin{equation}
\forall x,x' \in X, x \neq x' \Rightarrow f(x) \neq f(x').
\label{bijec}
\end{equation}
\item
All the ions and electrons present in the target after laser irradiation come
from the target before irradiation; i.e., onto mapping $=$ surjection,
\begin{equation}
f(X)=Y.
\label{srjec}
\end{equation}
\end{enumerate}
Therefore, the laser particle acceleration process is understood
as a bijective, that is injective and surjective, mapping $f$.

In topology, if the mapping $f$ is bijection and continuous,
and $f^{-1}$ is also continuous,
$X$ and $Y$ are called topologically isomorphic (homeomorphic)
and are denoted by $X \approx Y$.
$f$ is a continuous map, which means that the mapping $f$ is continuous at
each point of $X$. $f$ is continuous at point $x_0$ is written as
\begin{equation}
\forall \epsilon >0, \exists\delta >0 \mid d(x,x_0)<\delta \Rightarrow d(f(x),f(x_0))<\epsilon,
\label{conti}
\end{equation}
where $d(x, x_0)$ is the distance between $x$ and $x_0$.
This implies that $f$ does not leap at $x_0$ with a slight $x$ change.
It means that any continuous curve is transferred to a continuous curve.
Or, it can be defined using $\epsilon$-neighborhood at point $x_0$ of $X$,
$B(x_0;\epsilon)=\{x \in X \mid d(x,x_0) < \epsilon\}$.
If
\begin{equation}
\forall \epsilon >0, \exists\delta >0 \mid f(B(x_0;\delta)) \subset B(f(x_0);\epsilon)
\label{neigh}
\end{equation}
is satisfied at point $x_0 \in X$, $f$ is continuous at $x_0$.
This implies that the mapping,
which transfers nearby objects to the ones that are still nearby
without cutting or overlapping, is homeomorphism.

Figures $A, B$, and $C$ in Fig.\ref{fig:fig-mp}(b) are homeomorphic,
$A\approx B\approx C$.
On the other hand, figures $D, E$, and $F$ are non-homeomorphic
to $A$ and non-homeomorphic to each other,
$A \not\approx D \not\approx E \not\approx F$.

In this paper, we recognize the target deforming in the acceleration process
as shape changes, i.e., we see the target as a figure in the topology.
The correspondence from the target shape before deformation (the initial target)
to the deformed shape by laser irradiation (the deformed target)
is regarded as a mapping.
Determining whether two shapes are homeomorphic is a basic problem of topology.
In this paper,
we consider laser particle acceleration by focusing on whether the target
shapes before and after laser irradiation is homeomorphic.

In the acceleration process,
when an ion does not pass among the surrounding ions and its positional
relationship with the surrounding ions does not differ,
i.e., if an ion near other ions in the initial target is still located near
those ions after laser irradiation, the target deformation is a continuous map
and the targets before and after laser irradiation are homeomorphic, 
$X \approx Y$.
However, they are not homeomorphic, $X \not\approx Y$, in the acceleration
process such that some ions pass through other area ions.
Therefore, it is possible to see that the acceleration process
is different $\Rightarrow$ it is different in topology,
i.e., it is possible to know and understand the difference of the
acceleration scheme from the difference of the topology of the target.
Additionally, ions that are located in a narrow area, $U(x;\delta)$,
within the initial target, become ions with almost the same energy, i.e.,
the number and energy of the obtained ions are considered to be functions of
the coordinate $x$ in the initial target, and by studying its mapping, it is
possible to acquire in-depth understanding of the laser acceleration phenomenon.

\section{Simulation parameters} \label{h-para}

The simulations were performed with a parallelized electromagnetic code
based on the PIC method \cite{CBL}.
The parameters used in the simulations of the hydrogen foil target are
shown below.
An idealized model, in which a Gaussian linear polarized laser pulse is
normally incident on a foil target represented by a collisionless plasma,
is used.
The electron density, i.e., proton density,
is $n_{e}=5\times 10^{22}$ cm$^{-3}$.
The foil target has a thickness $1.0, 2.0,$ and $3.0 \mu$m in each case.
The total number of quasiparticles is $5\times 10^{8}$ in the case of
$1.0 \mu$m thickness, and it is two times for the $2.0 \mu$m thickness case
and three times for the $3.0 \mu$m case.
The number of grid cells is equal to $24000 \times 24000$ along
the $X$ and $Y$ axes, respectively.
Correspondingly, the simulation box size is $112\mu$m$\times 112\mu$m.
The laser propagation direction is set to be along the $X$ direction,
and the electric field is oriented in the $Y$ direction.
The boundary conditions for particles and fields are periodic in transverse
$Y$ directions and
absorbing at the boundaries of the computation box along the $X$ axis.
The laser-irradiated side surface of the foil is placed at $X=44 \mu$m,
and the center of the laser pulse is located $20 \mu$m behind it.
A $xy-$coordinate system is used throughout the text and figures.
The origin of the coordinate system is located at the center of
the laser-irradiated surface of the initial target,
and the directions of the $x$ and $y$ axes are same as
those of the $X$ and $Y$ axes, respectively.
Therefore,
the $x$ axis denotes the direction perpendicular to the target surface,
and the $y$ axes lie in parallel to the target surface.

We are particularly interested in the
intensity $I \sim 1\times 10^{22}$ W/cm$^{2}$, laser power $P \sim 1$ PW,
pulse duration $\sim 30$ fs full width at half maximum (FWHM),
and laser energy $ \mathcal{E}_\mathrm{las}\sim 20$ J,
which has been achieved recently in compact lasers \cite{Kiri}.
Therefore, the intensity is varied around $I=1\times 10^{22}$ W/cm$^{2}$.
Herein, the shape of the laser pulse, which is determined by pulse duration
and spot size, is the same in all cases.
When the peak intensity is $I$, the distribution of the laser intensity
$I^*$ is represented by $I^*(y,z,t)=I \cdot \psi (y, z, t)$.
We changed $I$, although the function $\psi(y, z, t)$
which indicates the shape of laser pulses is the same in all cases.
The laser energy is
\begin{equation}
 \mathcal{E}_\mathrm{las} = \int_{-\infty}^{\infty}\int_{-\infty}^{\infty}
 \int_{-\infty}^{\infty} I^*(y,z,t)dydzdt \nonumber \\
 = I\int_{-\infty}^{\infty}\int_{-\infty}^{\infty}\int_{-\infty}^{\infty}
 \psi(y,z,t)dydzdt=I\cdot V^*,
\label{els}
\end{equation}
where $V^*=\int_{-\infty}^{\infty}\int_{-\infty}^{\infty}\int_{-\infty}^{\infty} \psi(y,z,t)dydzdt$.
$V^*$ is the ``volume'' of the distribution function $\psi$ in the $y, z, t$
space and is the same in all cases under our conditions.
Therefore, the laser energy, $\mathcal{E}_\mathrm{las}$,
changes linearly with changes in laser intensity, $I$.
In the following, if there is an expression of change
in the laser intensity, it also implies change in laser energy.
Moreover, the same is true for laser power.
In this paper, we show a topological consideration for the target deformation,
especially noticing the deformation in the target thickness direction.
Therefore, in the simulation,
the target needs to be sufficiently finely divided in term of thickness.
We use a foil target consisting of hydrogen, since foil is the simplest,
and hydrogen is the simplest and has the highest possibility of generating
high energy ion \cite{TM1}.
Foil thicknesses take $1-3\mu$m, which is possible to generate
high energy protons in the above laser conditions \cite{TM1}.
Therefore, $1 \mu$m length, which is the thinnest thickness,
must be sufficiently finely divided for the simulation.
Also, a large simulation area is required,
since hydrogen occur a strong Coulomb explosion and is distributed
in a wide area.
That is, the number of grid cells needs to be very large.
Therefore, we performed two-dimensional (2D) simulations.

The laser pulse has $30$ fs in duration and focused on a spot size of
$2.5 \mu$m (FWHM) in all cases.
The laser peak intensity is variable,
$I=6.4 \times 10^{19} - 2.3 \times 10^{22}$ W/cm$^2$
($\mathcal{E}_\mathrm{las} = 0.1-53$ J).
At $I=1\times 10^{22}$ W/cm$^{2}$,
the laser peak power, $P$, is $783$ TW and
the laser energy, $\mathcal{E}_\mathrm{las}$, is $25$ J,
and the laser pulse with dimensionless amplitude
$a_0=q_eE/m_{e}\omega c=72$.

\section{Simulation results} \label{h-resu}

Simulation results using a hydrogen foil target are shown.
The change of the obtained proton energy and spatial distribution of protons,
electrons, are described by the intensity change,
i.e., energy and power also change,
upon a fixed laser pulse shape, i.e., spot size and pulse duration.
Three different foil thickness ware analyzed.

\begin{figure}[tbp]
\includegraphics[clip,width=8.0cm,bb=0 2 418 479]{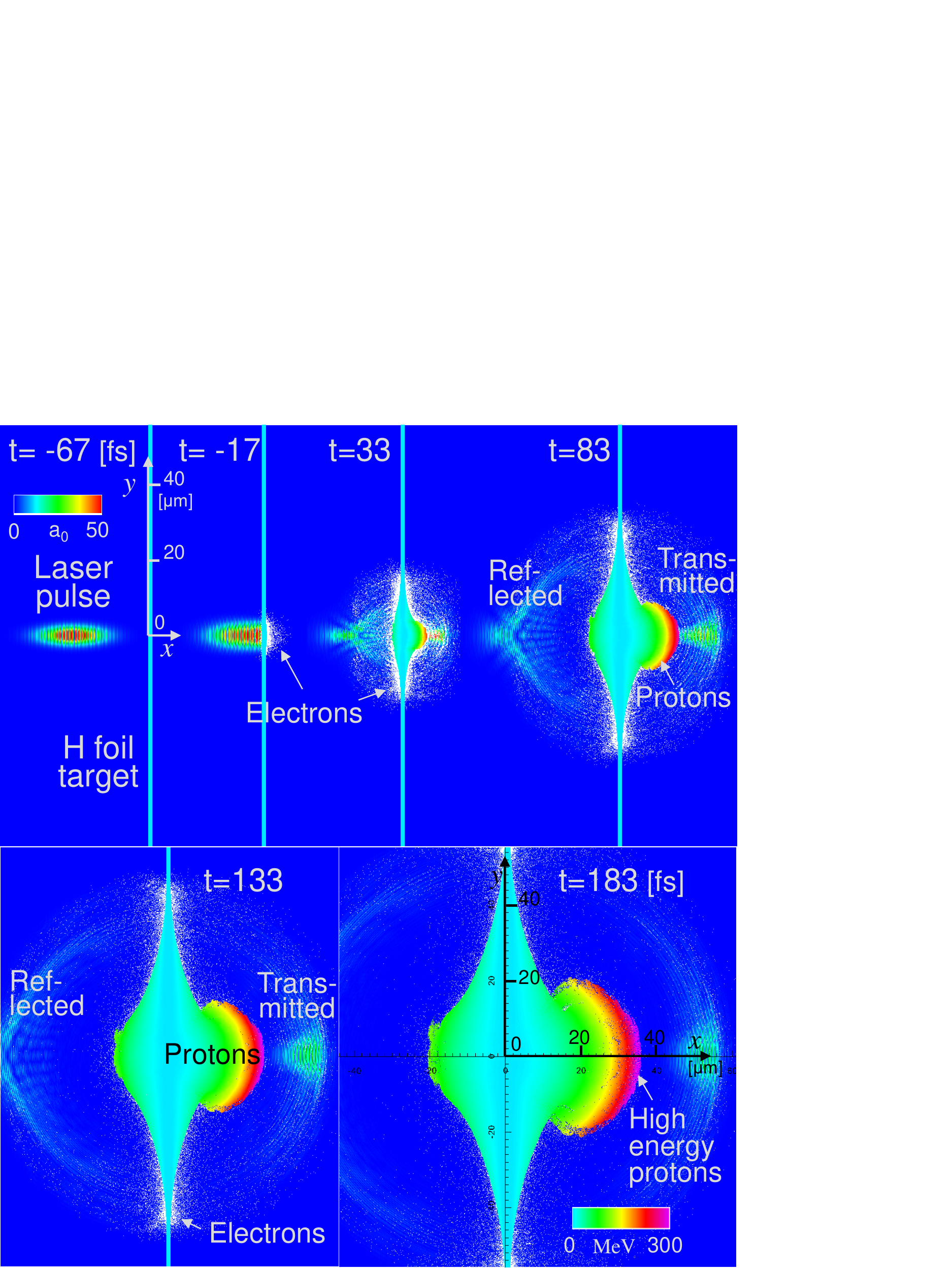}
\caption{
Spatial distribution of particles (protons and electrons) and
electric field magnitude
when $I=1 \times 10^{22}$ W/cm$^2$ ($\mathcal{E}_\mathrm{las}=25$ J)
and the target thickness $\ell=1\mu$m.
The time when the center of the laser pulse
reaches the surface$^-$ of the target is assumed to be $t = 0$.
For protons, the color corresponds to their energy.
}
\label{fig:fig-te}
\end{figure}

Figure \ref{fig:fig-te} shows
the particle distribution and the electric field magnitude
when a laser pulse of
$I=1 \times 10^{22}$ W/cm$^2$, $\mathcal{E}_\mathrm{las}=25$ J,
is normally incident on the hydrogen foil target of thickness $\ell = 1 \mu$m.
This is the laser parameter we are most interested in,
given the fact this intensity and energy are currently realized and
used in actual applications development.
Here, the time when the center of the laser pulse, where $I$ is the strongest,
reaches the position of the laser-irradiated surface of the initial target is
assumed to be $t = 0$.
This is because, the position of the laser pulse can be immediately recognized
from the time $t$,
since we consider while laser pulse and target are interacting in this paper.
That is, more than half of the laser pulse remains without interaction with
the target when $t<0$, and more than half of interaction have been done
at $t \geq 0$.
The simulation start time is $t=-67$ fs.
The initial shape of the laser pulse and the target is shown at $t=-67$ fs
in Fig. \ref{fig:fig-te}.
The laser pulse is defined on the $-x$ side of the target,
and it propagates in the $+x$ direction.
Hereinafter, the laser-irradiated surface of the target, $-x$ side surface,
is referred to as surface$^-$ and the opposite side surface,
the $+x$ side surface, is called surface$^+$.
At $t=-17$ fs, the head part of the laser pulse interacts with the target,
and mainly the electrons are pushed out in the $+x$ direction.
At this time, the center of the laser pulse is located at $x=-5 \mu$m,
which is behind the surface$^-$, and many parts of the laser pulse have
still not interacted with the target.
At $t=33$ fs, part of the laser pulse is reflected by the target while
another part is transmitted the target.
If there is no interaction with the target, the center position of the laser
pulse is at $x=10 \mu$m, and it has already passed the initial target position.
Therefore,
the strong interaction between the laser and the target has almost finished.
As $t=83, 133, 183$ fs, the target expands greatly with time.
The maximum proton energy at $t=183$ fs is $330$ MeV, and these appear on
the outermost part of the $+x$ side of the expanded target.
Proton energy gradually decreases as it goes from there in the $-x$ direction
up to $x>0$.
The distribution shape of protons is spherical in $+x$ side area and
mountainous-shaped in $-x$ side area.
The farthest position of the proton on the $+x$ side is $36 \mu$m, and
on the $-x$ side is $-20 \mu$m.
The $+x$ side expansion is 1.8 times longer than that of the $-x$ side,
i.e., the target expands greatly in the laser propagating direction.
Note that it is necessary to pay attention to the fact that the results shown
in this paper are for 2D simulation, and the obtained ion energies are evaluated
higher than in 3D simulation \cite{TM3}.

\begin{figure}[tbp]
\includegraphics[clip,width=8.0cm,bb=1 2 530 617]{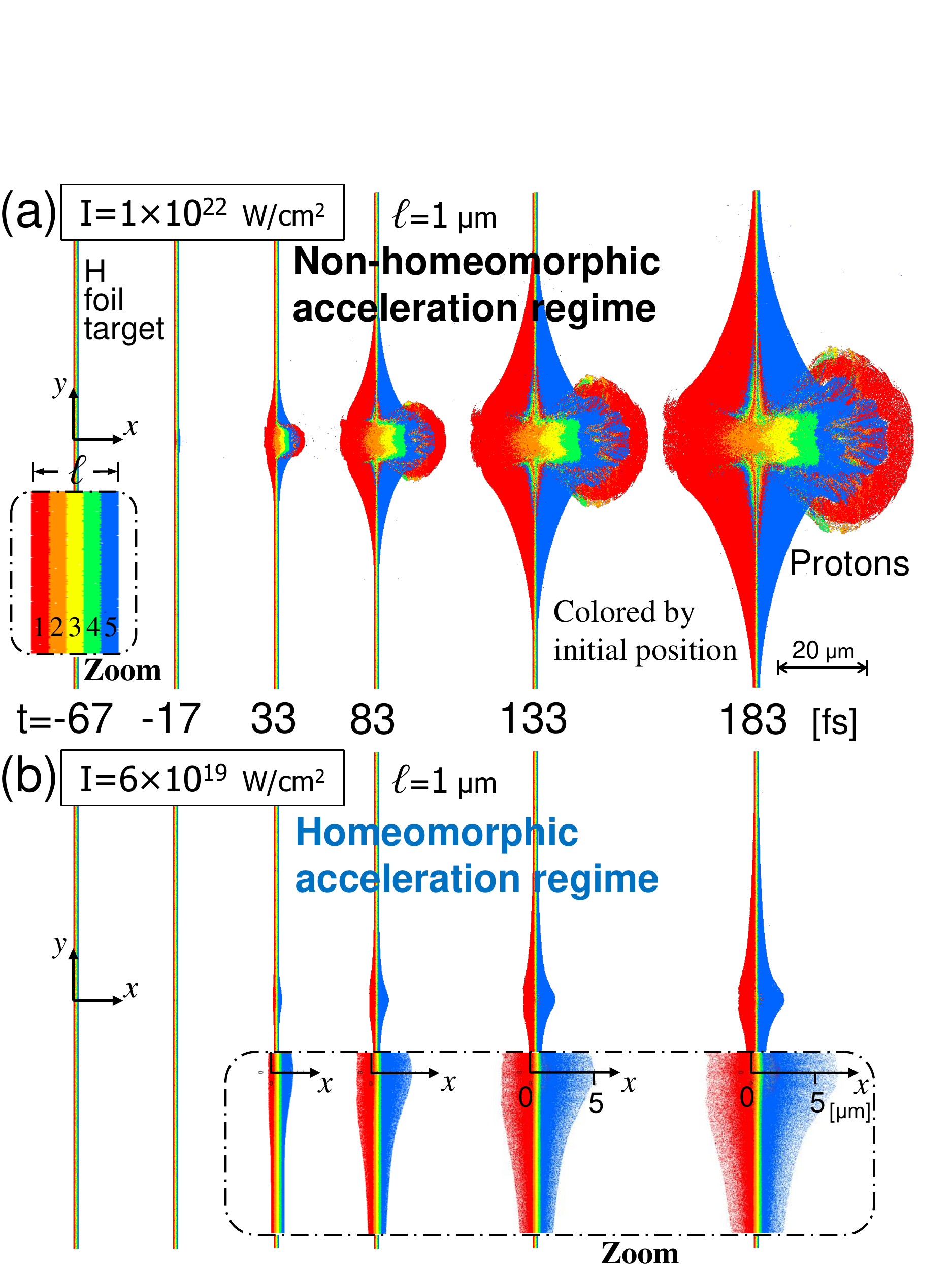}
\caption{
Spatial distribution of protons in the cases of
$I=1 \times 10^{22}$  W/cm$^2$ ($\mathcal{E}_\mathrm{las}=25$ J) (a),
and $I=6 \times 10^{19}$ W/cm$^2$ ($\mathcal{E}_\mathrm{las}=0.1$ J) (b).
The color corresponds to their initial positions.
The initial region is determined by slicing the foil into five regions
(see zoomed figure at $t=0$ in (a)),
and each region is numbered with the number shown in the figure.
(a) The ion distributions are non-homeomorphic,
and (b) these are homeomorphic.
}
\label{fig:fig-ti}
\end{figure}

In Fig. \ref{fig:fig-ti}(a),
the same result as in Fig. \ref{fig:fig-te} is shown with a different way.
However, only protons are shown here.
Protons are color-coded by their initial position
(see zoomed target at $t=-67$ fs in Fig. \ref{fig:fig-ti}(a)).
The initial foil is sliced evenly into five in the thickness direction,
$x$ direction, and each region of protons is of a different color.
This makes it possible to know where the protons in a deformed target come
from on the initial target.
Each region is called region-1, region-2,.., region-5, sequentially from
the $-x$ side.
As shown at the final time $t=183$ fs,
we can see that the protons near the $+x$ side outermost part come from
near the opposite side, surface$^-$.
That is, the target before laser irradiation and
the target after irradiation are non-homeomorphic.
Furthermore, by looking at this together with Fig. \ref{fig:fig-te},
it can be seen that the high energy protons are the protons near the
laser-irradiated surface, surface$^-$.
Also, at an early time, $t=33$ fs,
the protons near the surface$^-$ have already appeared on the $+x$ side target
outermost part, indicating that its distribution is formed at the early stage
of the acceleration process.

Figure \ref{fig:fig-ti}(b) shows the result of $I=6 \times 10^{19}$ W/cm$^2$,
$\mathcal{E}_\mathrm{las}=0.1$ J, in the same way as
in Fig. \ref{fig:fig-ti}(a).
The times are also the same as in Fig. \ref{fig:fig-ti}(a).
The figures in the dashed-and-dotted frame are enlarged views including
the center of the target.
At weak laser intensity, i.e., power and energy,
the protons after laser irradiation do not distribute like those
in Fig. \ref{fig:fig-ti}(a).
During the acceleration process
each region is distributed in the order of the initial target at all times,
without the protons near the surface$^-$
going out to the region near surface$^+$.
That is, both the target before laser irradiation and the target
after irradiation are homeomorphic.
And, in Figs. \ref{fig:fig-ti}(a) and \ref{fig:fig-ti}(b) with different
laser intensities, energies,
the topology of the target after laser irradiation is different.
Here, we define the terms used in this paper.
When the deformed target by laser irradiation is homeomorphic
with the initial target,
we call this acceleration `homeomorphic acceleration' (HA).
Since the initial target and the targets after laser irradiation in
Fig. \ref{fig:fig-ti}(b) are homeomorphic, this is HA.
On the contrary,
when the deformed target by laser irradiation is non-homeomorphic with
the initial target, as shown in Fig. \ref{fig:fig-ti}(a),
we call that acceleration `non-homeomorphic acceleration' (NHA).
Also, when the target after deformation is homeomorphic with the initial target,
the distribution after acceleration is called homeomorphic
distribution (H distribution).
When it is non-homeomorphic,
this is called non-homeomorphic distribution (NH distribution).
When the positions of the ions (protons) change vigorously and
the ions (protons) of region-1, the red part,
are widely distributed in the $+x$ side farthest area of the target,
like $t=183$ fs in Fig. \ref{fig:fig-ti}(a), we call this strong NHA.
When it is NHA, but not change as vigorously, we call this weak NHA.
In addition, at $t=183$ fs in Fig. \ref{fig:fig-ti}(b),
looking closely at the center part of the target around $x=y=0$,
the protons near the surface$^-$ are slightly existed in the region-5 protons,
where around $x=2 \mu$m, although the number is very small.
We look at this from a wider point of view and express it as HA.
Depending on the degree, it is expressed as almost HA.

\begin{figure}[tbp]
\includegraphics[clip,width=8.0cm,bb=5 0 404 244]{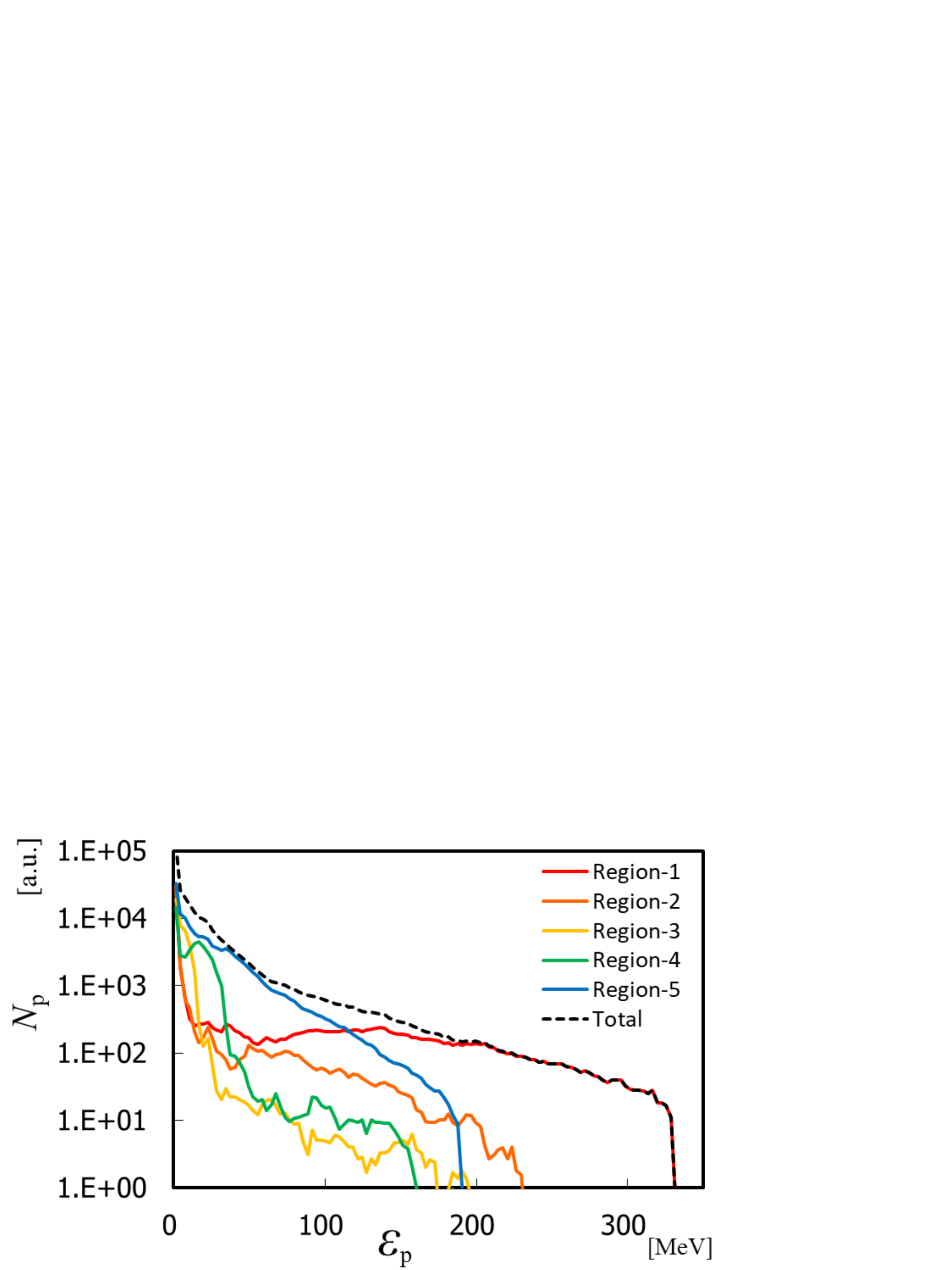}
\caption{
The energy spectrum of the proton beam in the case of
Fig. \ref{fig:fig-te} at $t=183$ fs.
Hydrogens of region-1,
presenting a laser irradiation surface,
become the highest energy, 
and next highest energy region is region-2.
}
\label{fig:fig-es}
\end{figure}

Figure \ref{fig:fig-es} shows the energy spectra of the protons at $t=183$ fs
in the case shown in Fig. \ref{fig:fig-te} (also Fig. \ref{fig:fig-ti}(a)).
Since the protons that we use for some applications are protons accelerated
in the $+x$ direction, i.e., laser propagated direction,
here we show only the result of protons accelerated in the $+x$ direction.
The energy spectra for all the protons, represented by a dotted line, and each
region of protons by a solid color-coded line for each region, are shown.
All the high energy protons are from region-1 which has the laser
irradiation surface, i.e., surface$^-$, and next highest energy region is
region-2 which is the next region to it.
The difference between the maximum proton energies of these two regions
is quite large, at about $100$ MeV.
That is, the high energy protons all come from near surface$^-$.
The maximum proton energy of region-1 is prominently high.
The differences in the maximum proton energies of regions 2--5 is
not very different in comparison with that between them and in region-1.

\begin{figure}[tbp]
\includegraphics[clip,width=8.0cm,bb=20 0 420 379]{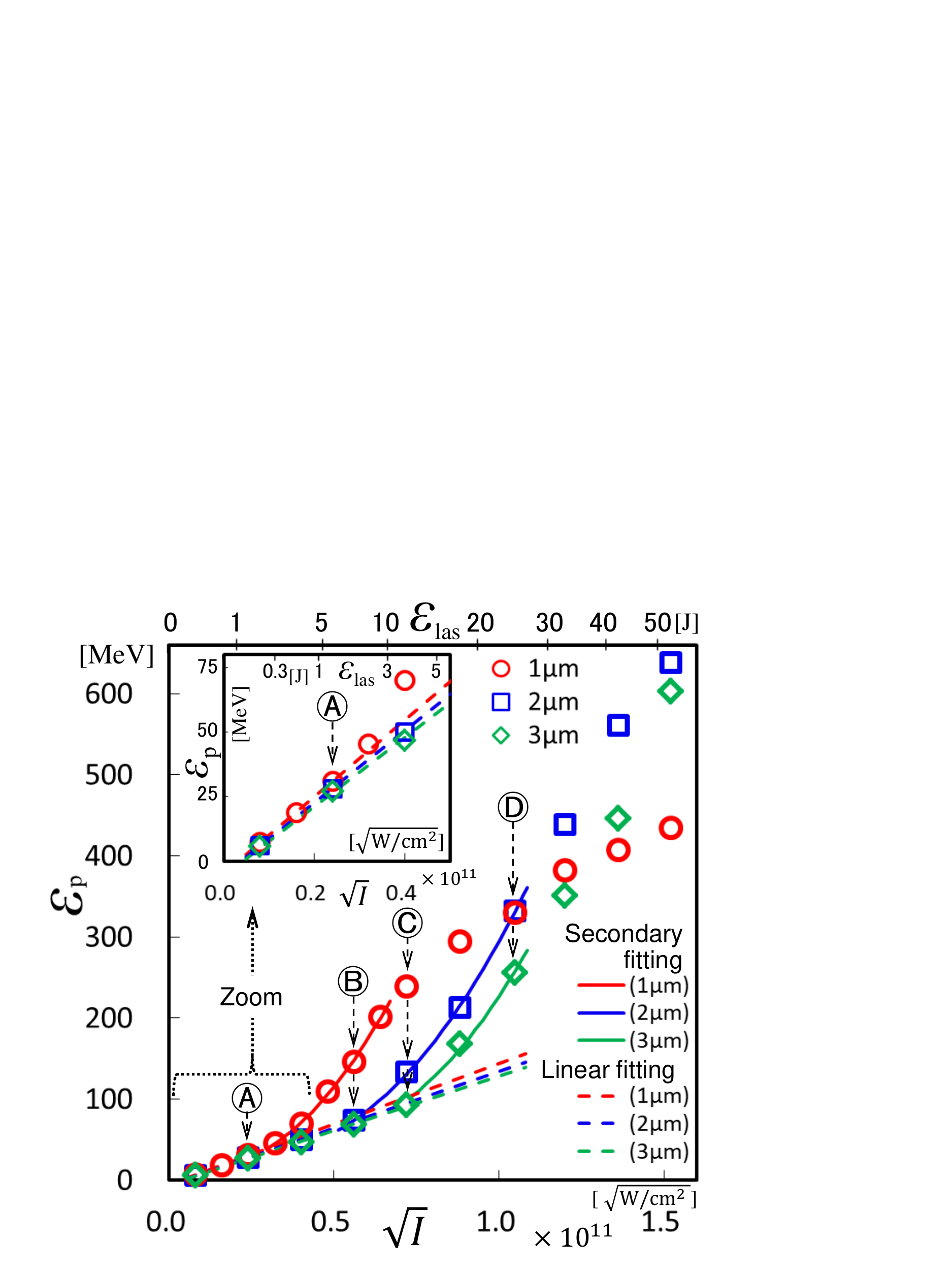}
\caption{
Maximum proton energy, $\mathcal{E}_{p}$, in each foil thickness
as a function of square root of the laser intensity, $\sqrt{I}$,
(also $\sqrt{\mathcal{E}_\mathrm{las}}$).
The maximum proton energy rises in proportion to
$\sqrt{I}$, $\sqrt{\mathcal{E}_\mathrm{las}}$, and is not greatly related to
the thickness in low $I$. In the region above a certain $I$,
it rises in the second-order to $\sqrt{I}$, $\sqrt{\mathcal{E}_\mathrm{las}}$,
and greatly affects the thickness.
}
\label{fig:fig-ip}
\end{figure}

Figure \ref{fig:fig-ip} shows the maximum proton energy at each laser intensity
$I$, and also laser energy $\mathcal{E}_\mathrm{las}$.
The results of three foil thickness cases of $1, 2, 3 \mu$m are shown.
The laser intensity, $I$, on the bottom horizontal axis is indicated by
its square root, $\sqrt{ I}$.
The upper horizontal axis shows the laser energy, $\mathcal{E}_\mathrm{las}$.
The inset shows the zoom of
$0 \leq \sqrt{I} \leq 0.5 \times 10^{11} \sqrt{\mathrm{W/cm}^2}$.
The stronger $I$, $\mathcal{E}_\mathrm{las}$,
the higher the energy of the protons are generated.
At low $I$,  $\mathcal{E}_\mathrm{las}$, the maximum proton energy rises in
proportion to $\sqrt{I}$, $\sqrt{\mathcal{E}_\mathrm{las}}$.
The dotted lines are the linear fitting of the part where the proton energy
is proportional to $\sqrt{I}$.
The points used for the linear fitting are the values of
$\sqrt{I} \leq 0.24 \times 10^{11} \sqrt{\mathrm{W/cm}^2}$ for
$1 \mu$m thickness,
$\sqrt{I} \leq 0.56 \times 10^{11} \sqrt{\mathrm{W/cm}^2}$ for
$2 \mu$m thickness,
and $\sqrt{I} \leq 0.72 \times 10^{11} \sqrt{\mathrm{W/cm}^2}$ for
$3 \mu$m thickness.
The proportionality continues with the stronger $I$
as the foil thickness increases.
Moreover, the obtained proton energy is not greatly related to foil thickness
in the proportional region.
The proton energy in this region is almost same at each thickness,
although the foil thickness is up to three times different.
For example,
at a point of intensity indicated by $\textcircled{\scriptsize A}$,
$\sqrt{I}=0.24 \times 10^{11} \sqrt{\mathrm{W/cm}^2}$, in the figure,
the proton energies, which are generated from $1, 2$ and $3 \mu$m thickness are
$31, 28$, and $27$ MeV, respectively, and are approximately the same.
As $\sqrt{I}$ increases, the rise of proton energy changes from
a linear to a second-order steep rise curve.
The solid lines in the figure are the second-order fitting lines with
the proton energy.
The proton energy rises in second-order with $\sqrt{I}$, that is,
linearly with $I$.
This second-order rise starts earlier with a thinner thickness of foil.
In this region, the obtained proton energy differs greatly depending on
the difference in foil thickness.
For example, in the proton energy at point $\textcircled{\scriptsize C}$,
the thickness of $1$ and $2 \mu$m are in the second-order increasing region
and the thickness of $3 \mu$m is in the linearly increasing region.
The proton energies are $240, 133, 93$ MeV for $1, 2$, and $3 \mu$m,
respectively. The maximum difference is $2.6$ times.
It can be said that,
the obtained proton energy greatly different due to the
difference in the target thickness in the region above a certain intensity.
On the other hand,
the target thickness does not significantly influence
the obtained proton energy in the region of low laser intensity.

\begin{figure}[tbp]
\includegraphics[clip,width=8.0cm,bb=14 1 333 525]{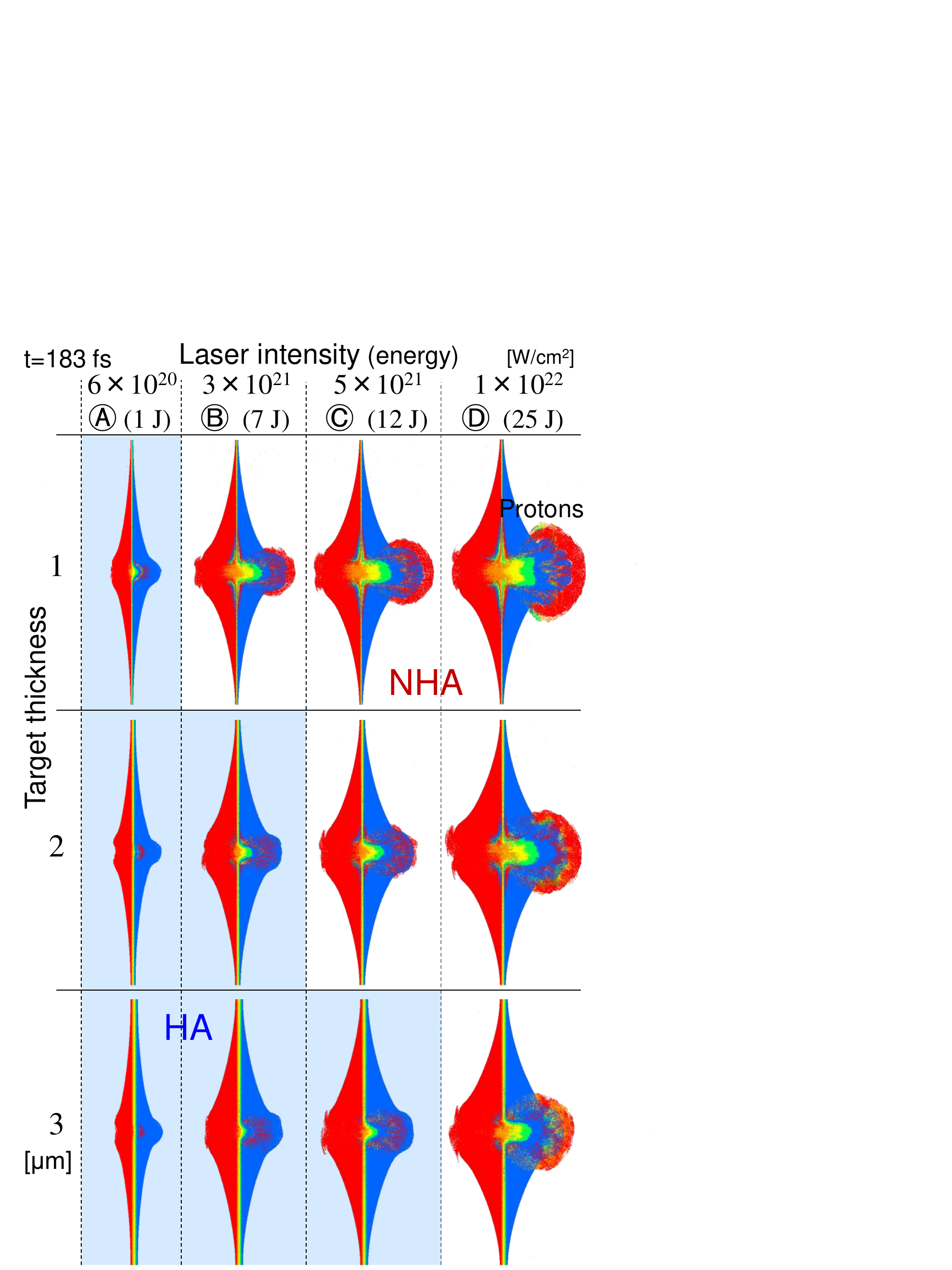}
\caption{
The distributions of protons for each target thickness and
for each laser intensity (energy).
The protons are color-coded by their initial positions.
The white parts of the background are NHA,
and the light blue parts are HA or almost HA.
Where the proton energy rises in the second-order of $\sqrt{I}$,
$\sqrt{\mathcal{E}_\mathrm{las}}$, they are the NHA regime.
}
\label{fig:fig-ii}
\end{figure}

The distributions of protons at $t = 183$ fs for each target thickness of
$1, 2, 3 \mu$m and for the laser intensity of each
$\textcircled{\scriptsize A}$, $\textcircled{\scriptsize B}$,
$\textcircled{\scriptsize C}$, and $\textcircled{\scriptsize D}$,
are shown in Fig. \ref{fig:fig-ii}.
The protons are color-coded by their initial positions,
as in Fig. \ref{fig:fig-ti}.
The $1 \mu$m thickness case of the laser intensity $\textcircled{\scriptsize D}$
is the same as $t=183$ fs in Fig. \ref{fig:fig-ti}(a).
These are almost HA at all thicknesses in the laser
$\textcircled{\scriptsize A}$,
and are all strong NHA in the laser $\textcircled{\scriptsize D}$.
We can see that the points where proton energy raise in a linear manner
with $I$ (second-order with $\sqrt{I}$),
i.e., $1 \mu$m-$\textcircled{\scriptsize B}$ $\textcircled{\scriptsize C}$,
$2 \mu$m-$\textcircled{\scriptsize C}$ $\textcircled{\scriptsize D}$, and
$3 \mu$m-$\textcircled{\scriptsize D}$,
are strong NHA, by looking at it together with Fig. \ref{fig:fig-ip}.
That is, it can be said that the obtained proton energy rapidly rises in
proportion to the laser intensity $I$ (laser energy $\mathcal{E}_\mathrm{las}$)
when a strong NHA occurs.
On the other hand,
where the obtained prototype energy is proportional to
$\sqrt{I}$. i.e., $1 \mu$m-$\textcircled{\scriptsize A}$,
$2 \mu$m-$\textcircled{\scriptsize A}$ $\textcircled{\scriptsize B}$, and
$3 \mu$m-$\textcircled{\scriptsize A}$ $\textcircled{\scriptsize B}$
$\textcircled{\scriptsize C}$,
these are almost HA.

The proton energy rises in the second-order of $\sqrt{I}$,
$\sqrt{\mathcal{E}_\mathrm{las}}$, in the NHA regime,
and rises in proportion to $\sqrt{I}$, $\sqrt{\mathcal{E}_\mathrm{las}}$,
in the HA regime.
Moreover, the target thickness strongly affects
the obtained proton energy in the NHA regime,
and the obtained proton energy is not significantly related to the target
thickness in the HA regime.
From another point of view, we consider the change in proton energy when
the target thickness is gradually reduced at a certain laser intensity, energy.
When the target is thick, it is the HA regime,
and even if the thickness is changed making it slightly thinner,
the obtained proton energy does not change much up to a certain thickness.
However, when it is less than a certain thickness, it becomes the NHA regime,
the produced ion energy becomes rapidly higher,
and the ion energy becomes higher as the thickness becomes thinner.

The NHA regime is considered in detail below.
The case of thickness $\ell=1 \mu$m, and $I=1 \times 10^{22}$ W/cm$^2$,
$\mathcal{E}_\mathrm{las} =25$ J, which is the
$\textcircled{\scriptsize D}$ case laser, is used for the considerations.
The $1 \mu$m thickness is selected because it generates higher energy protons
than the other thicknesses at laser intensities of
$\textcircled{\scriptsize A} - \textcircled{\scriptsize D}$.
The selection of $I=1 \times 10^{22}$ W/cm$^2$ is due to the fact that
it is NHA for all $1, 2$, and $3 \mu$m thickness cases and
this intensity has recently been achieved.
As shown in Fig. \ref{fig:fig-ti}(a),
we see that the NH distribution has already appeared,
i.e., the formation processes has ended, at the initial time $t=33$ fs.
Therefore, we investigate in detail the movement and location of electrons and
ions at the initial time, $t < 0$ fs.
Below, since the contrast between negatively charged particles, i.e., electrons,
and positively charged particles, i.e., ions, in the target are important,
we generally refer to as ions without writing protons.
The laser acceleration phenomenon is shown in the time interval during which
the laser pulse and the target interact, that is,
in the time about the laser pulse duration.
Since the pulse width $=30$ fs,
the results over a very short time are shown below.

\begin{figure}[tbp]
\includegraphics[clip,width=7.5cm,bb=2 3 388 315]{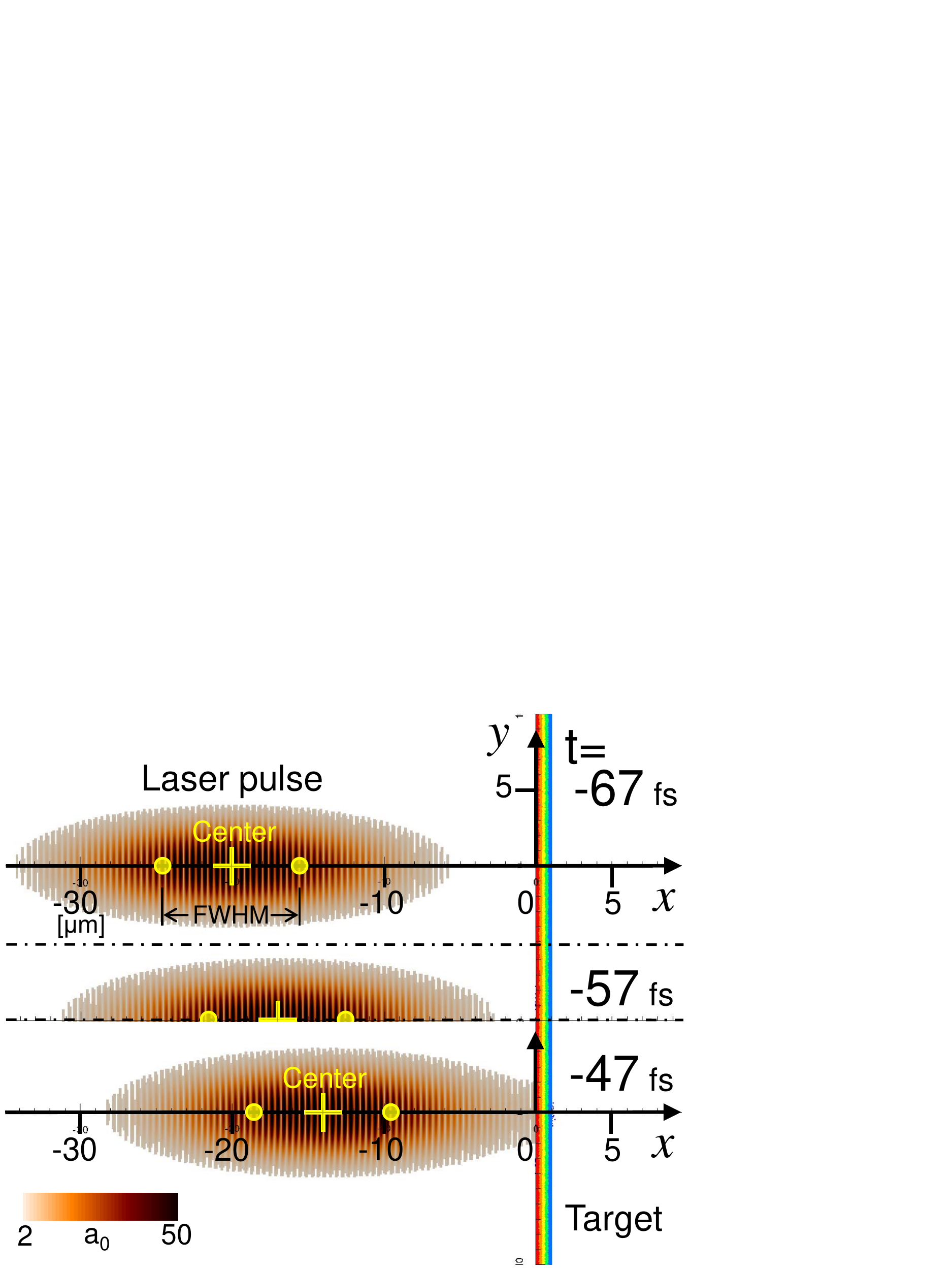}
\caption{
The target and the laser pulse (isosurface for value $a_0=2$)
of the $I=1 \times 10^{22}$ W/cm$^2$ ($\mathcal{E}_\mathrm{las}=25$ J),
$\ell=1 \mu$m case at early simulation times $t=-67$, $-57$, and $-47$ fs.
The center and FWHM points of the laser pulse are indicated by
'$+$' and '$\bigcirc$', respectively.
The start time of the interaction between the laser and
the target is approximately $t=-47$ fs.
}
\label{fig:fig-ini1}
\end{figure}

Figure \ref{fig:fig-ini1} shows
the laser pulse and the target at time $t=-67,-57$, and $-47$ fs.
It is shown by zooming the area near the initial laser pulse and
the part of the target where the laser irradiates. That is,
the $x$ direction is in the range of $x=-35$ to $10 \mu$m and the $y$
direction is in the range of $y \simeq -10$ to $10 \mu$m.
In this study, since the interaction between the target and
the laser pulse at the initial time is important,
the laser pulse is placed relatively far away from the target so that
there is no interaction at the simulation start time, $t=-67$ fs.
At $t=-57$ fs, the interaction between the laser pulse and the target
has scarcely begun to occur yet.
At $t=-47$ fs, the center of the laser pulse is at $x=-14 \mu$m and
FWHM position on the side closer to the target is at $x=-9.5 \mu$m.
The distance from this FWHM position to the surface$^-$ is also about FWHM.
At this time,
the interaction between the laser pulse and the target has started slightly.
Therefore, it is assumed that the time when
the interaction starts between the laser pulses and
the target is $t \approx -47$ fs, and in the flowing, we investigate within
approximately the pulse width $=30$ fs from this start time.
That is, the acceleration process from $t=-47$ to $-7$ fs, which
is the time just before the center of the laser pulse reaches
the target, is considered in detail.

\begin{figure}[tbp]
\includegraphics[clip,width=8.5cm,bb=14 10 504 392]{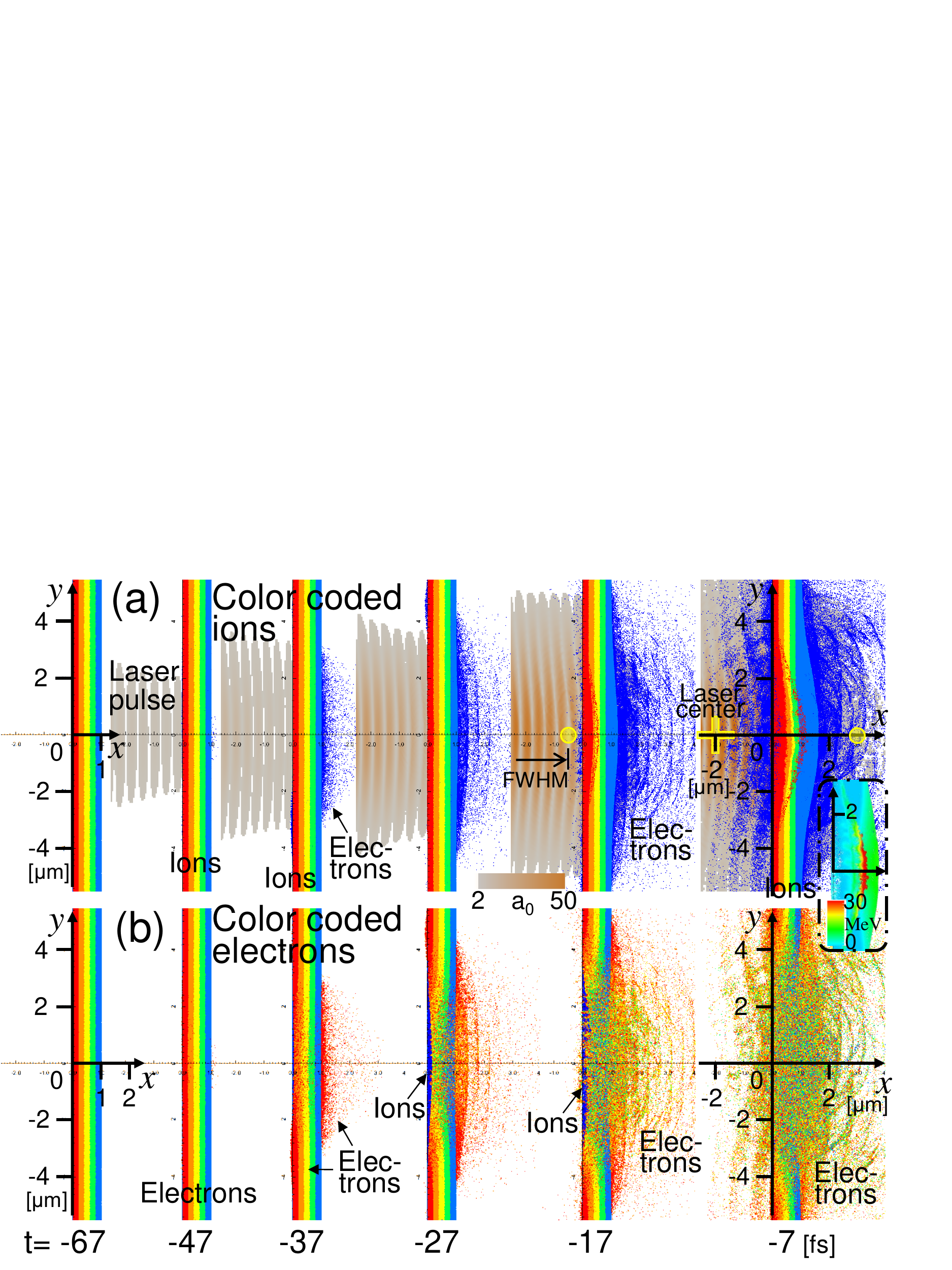}
\caption{
(a) Spatial distribution of electrons and ions (protons),
and electric field magnitude (isosurface for value $a_0=2$).
Ions are color-coded by their initial position.
(b) Electrons are color-coded.
In the inset: the color corresponds to the ions' energy.
}
\label{fig:fig-ini_ie}
\end{figure}

\begin{figure}[tbp]
\includegraphics[clip,width=7.5cm,bb=16 3 326 718]{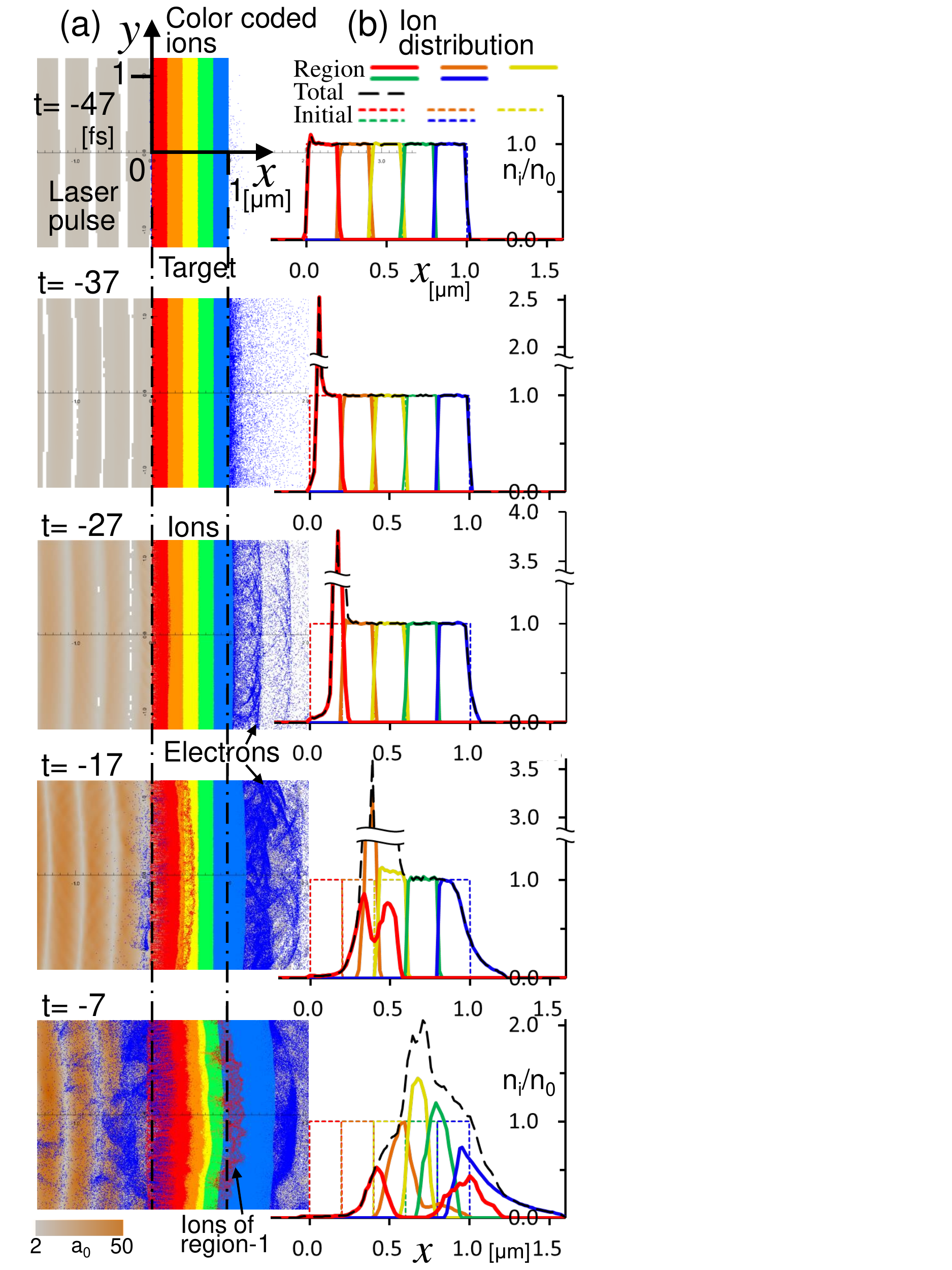}
\caption{
(a) Spatial distribution of electrons and ions (protons).
Ions are color-coded according to the region of their initial position.
(b) Distributions of the number of ions
for each initial region in the $x$ direction.
The number of ions per unit length along $x$, $n_{i}$,
is normalized by its initial value, $n_0$.
Some of the reagion-1 ions, red color ions, are passing through regions 2--4
and are going toward the $+x$ side, opposite side, surface.
}
\label{fig:fig-ini_i}
\end{figure}

\begin{figure}[tbp]
\includegraphics[clip,width=7.5cm,bb=7 3 315 718]{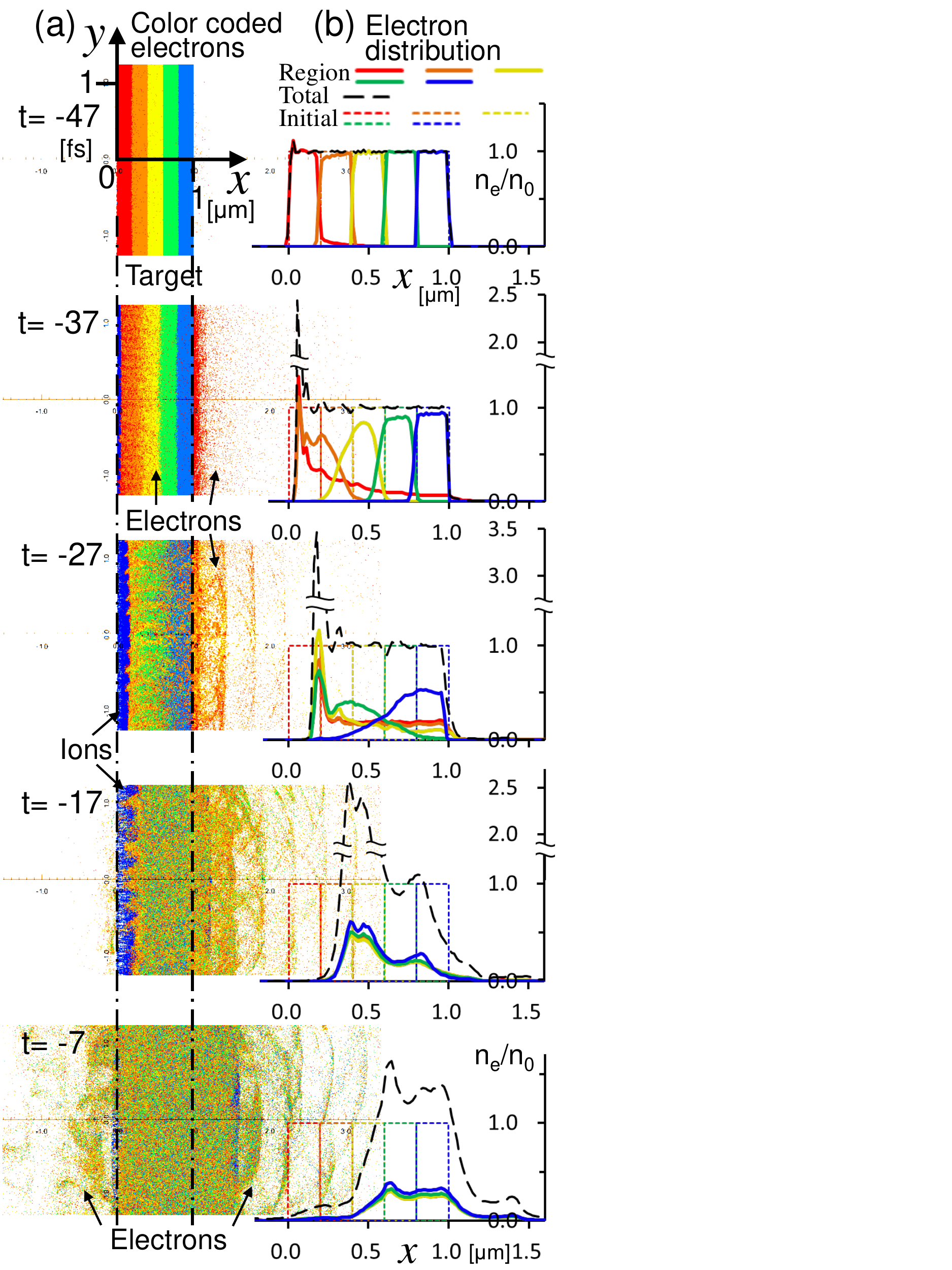}
\caption{
(a) Spatial distribution of electrons and ions.
Electrons are color-coded according to the region of their initial position.
(b) Distributions of the number of electrons
for each initial region in the $x$ direction.
The number of electrons per unit length along $x$, $n_{e}$,
is normalized by its initial value, $n_0$.
The electrons near the laser irradiation surface are pushed away
in the $+x$ direction, and the electrons present inside
the target move in the $-x$ direction and flow into there,
and then next these flowing electrons are pushed away.
}
\label{fig:fig-ini_e}
\end{figure}

Figures \ref{fig:fig-ini_ie}, \ref{fig:fig-ini_i}, and
\ref{fig:fig-ini_e} show the distributions of ions, i.e., protons,
and electrons at the initial time $t \leq -7$ fs.
An overview is shown schematically in Fig. \ref{fig:fig-ini_ie} and detailed
distributions are shown in Figs. \ref{fig:fig-ini_i}, \ref{fig:fig-ini_e}.

Figures \ref{fig:fig-ini_ie}(a) and (b) are the same results of the
distribution of the ions and the electrons, except for their colors.
Ions in Fig. \ref{fig:fig-ini_ie}(a) and electrons in
Fig. \ref{fig:fig-ini_ie}(b) are color-coded into
five colors, according to their initial positions.
The figures at $t=-67$ fs are in the initial state.
The laser pulse is also drawn in Fig. \ref{fig:fig-ini_ie}(a),
the center of it is indicated by '$+$', and the $+x$ side FWHM point is
indicated by a '$\bigcirc$'.
These figures are zoomed views around $x=y=0$,
the $y$ direction displayed area length is $10 \mu$m which is four times
the laser spot diameter.
First, electrons in the target are pushed out to the $+x$ side of the target
by the laser pulse, and then the ions expand gradually.
Let us first consider movement and distribution of the ions
in Fig. \ref{fig:fig-ini_ie}(a).
As shown at $t=-17$ fs,
when the $+x$ side FWHM point of the laser pulse reaches
the surface$^-$, some of the ions of region-1, red part,
have already come into region-3, the yellow area.
That is, NHA has already started at $t \sim -17$ fs.
At $t=-7$ fs, when the $+x$ side FWHM point has passed through the target and
the pulse center has not yet been reached, some of the ions of region-1
reach region-5, the light blue area.
That is NHA.
At this time,
region-5 expands greatly, and the other regions move in
the $+x$ direction while maintaining almost the initial thickness.
The inset in Fig. \ref{fig:fig-ini_ie} shows the ion distribution at $t=-7$ fs,
the color corresponds to their energy.
By looking at it together with $t=-7$ fs in Fig. \ref{fig:fig-ini_ie}(a),
we see that some of the ions of region-1, which separated from
the other region-1 ions, are positioned near the center of the target and
have higher energy than the other ions in the target.
The energy of these ions is around $40$ MeV,
and that of the other regions ions around them is around $10$ MeV.
That is, at this time, some of the region-1 ions form a high energy bunch,
and this moves and accelerates in the $+x$ direction
and passes through the other regions.

Next, the movement and distribution of the electrons are
considered (Fig. \ref{fig:fig-ini_ie}(b)).
The start time of the movement of the electrons is much earlier than that of
the ions,
and many electrons appear in the $+x$ side area of the target at $t=-37$ fs.
These electrons are of region-1, the red area.
The electrons near the surface$^-$ are also accelerated first,
in a similar way to the ions, and pass through the electrons in the other
regions and exiting on the $+x$ side of the target.
Thereafter,
the electrons of the other regions are also accelerated in the $+x$ direction,
and the electrons of each region are almost evenly mixed, at $t=-7$ fs.
The movements of the ions and electrons of each region
in the early simulation time are shown in more detail below.

First, we show the movement of ions, i.e., protons.
Figure \ref{fig:fig-ini_i} shows the spatial distributions of the ions and
electrons, and the distributions of the number of ions in the $x$ direction.
Figure \ref{fig:fig-ini_i}(a) shows the distribution of the ions,
color-coded by region, and the distribution of the electrons with
a single color, dark blue, and the laser pulse.
Although these are the same results as those in Fig. \ref{fig:fig-ini_ie}(a),
the area around the target center is zoomed more in here.
The times are $t =-47$ to $-7$ fs,
which are the same times as Fig. \ref{fig:fig-ini_ie}.
Figure \ref{fig:fig-ini_i}(b) shows the distributions of the number of ions
in the $x$ direction, and this is shown in different colors for each region.
Here, the ions in the range of $y=-1.25$ to $1.25 \mu$m are counted.
The number density of all the ions, which is not divided into regions, is
indicated by a dotted black line, and the initial distributions of the ions
of each region are indicated by a dotted line according to each color.
The density is shown as a ratio to the initial density, $n_i/n_0$.
The initial number density of the ions, $n_{i0}$, and that of the electrons,
$n_{e0}$, are equal in the hydrogen target, $n_{i0}=n_{e0}=n_0$.
Although there is still no great change from the initial state at $t=-47$ fs,
a slightly high-density area of region-1 ions
are generated near the surface$^-$ (Fig. \ref{fig:fig-ini_i}(b)).
At $t=-37$ fs, particularly noticeable high-density peaks appear
near the surface$^-$ in the ion distribution.
This peak of ion density of region-1 moves in the $+x$ direction over time.
Then, at $t=-27$ fs, this peak becomes even higher,
with the maximum point of density being $3.8$ times the initial density.
In the spatial distribution of ions (Fig. \ref{fig:fig-ini_i}(a)),
the boundaries of each ion region are clear in $t \le -27$ fs,
and the ions of each region are scarcely mixed.
At $t=-27$ fs, the surface$^-$ moves slightly in the $+x$ direction,
and the ions near the surface$^+$ expands slightly to the $+x$ side area
of the target.
At $t=-17$ fs, some of the ions of region-1 have entered region-3,
and the high-density peak of it begins to split into two.
At $t=-7$ fs, some of the region-1 ions pass through regions 2--4 and
reach the region-5, and region-1 splits into two parts.
The peak density of the region-2 ions appears at $t=-17$ fs
after the peak of the region-1 becomes smaller.
It is region-3 that forms the next peak to this (see $t=-7$ fs).
That is, the ions in the target form a high-density peak in order, from
the surface$^-$ side to the $+x$ side region, over time.
These peaks move in the $+x$ direction.
However, the ions of region-5 do not form a high-density peak like
the other regions, and expand greatly in the $+x$ direction over time.

Similar figures are shown for the electrons in Fig. \ref{fig:fig-ini_e}.
Figure \ref{fig:fig-ini_e}(a) shows the spatial distributions of the electrons,
color-coded by region,
and the distribution of the ions in a single color, namely dark blue.
Fig. \ref{fig:fig-ini_e}(b) shows
the distributions of the number of electrons in the $x$ direction.
The way to make these figures and display times
are the same as for the ion (Fig. \ref{fig:fig-ini_i}).
The distributions of the electrons of each region have already changed
at $t=-37$ fs,
although in contrast, the ion distributions maintain almost the initial
distributions at this time (see Fig. \ref{fig:fig-ini_i}).
At $t=-47$ fs, a slightly high-density area of electrons of region-1
appears near the surface$^-$, in a way to similar to the ions.
Looking at the boundary between region-1, the red line,
and region-2, the orange line, of the number density distribution,
the electrons of region-1 enter deeply into region-2,
and even reach region-3.
Some of the electrons of region-2
move a little in the $-x$ direction, and enter slightly into region-1.
That is, some of the electrons of region-1 move in the $+x$ direction, and
some of the electrons of region-2 move in the opposite, $-x$, direction.
This is explained as follows.
The laser pulse interacts first with the region-1 electrons which
are located closest to it.
In this strong interaction between the laser pulse and the region-1 electrons,
the electrons near to the surface$^-$ gain great momentum in the $+x$ direction
by the ponderomotive force of the laser pulse, and move in the $+x$ direction.
At this time, the ions which are much heavier than the electrons hardly move
and, as a result, the area where those electrons are located has a $+$charge.
The electrons of region-2 which have still almost no momentum,
and are nearly stationary, flow into this $+$charge area by Coulomb force.
At $t=-37$ fs in Fig. \ref{fig:fig-ini_e}(a), many electrons of region-1
appear on the $+x$ side in the outer area of the target.
Looking at the density distribution at this time (Fig. \ref{fig:fig-ini_e}(b)),
the $-x$ side surface of the electrons of region-1 move in the $+x$ direction,
and a peak of density appears at a position slightly
$+x$ side from this surface.
The region-1 electrons are distributed in the outer area on the $+x$ side of
the target, while reducing the number density in the $+x$ direction
from that peak.
On the other hand, in contrast, at this time many electrons of region-2
move from the initial position to the $-x$ direction,
and the main distribution area moves close to the initial region-1 area.
The peak of the number density is also located in the initial area of
region-1, and is at the same positon as that of the region-1 electrons.
In the vicinity of those peak positions,
the electrons of regions-1 and 2 are approximately equally mixed.
At this time,
the electrons near the $-x$ side boundaries of regions-3, 4, and 5 also move
in the $-x$ direction,
and the amount of movement is larger for the region on the $-x$ side.
The reason for the movement is the same as for the movement in
region-2 at $t=-47$ fs mentioned above,
i.e., Coulomb force on the $+$charge near the surface$^-$.
At $t=-27$ fs, the distribution areas of the electrons of
regions-1, 2, 3, and 4 are approximately the same,
and the peaks also appear at approximately the same positions.
That is, they are almost equally mixed.
The number density of the peak of all the electrons is about $3.5$ times
the initial density and is the highest of all the times.
On the other hand,
region-5 still has an independent distribution at this time.

To summarize the above, the electrons located near the laser irradiation area of
the target form a peak of electron density, i.e., high-density electron bunch,
by the ponderomotive force of the laser pulse,
and that peak gains momentum in the $+x$ direction and
moves in the $+x$ direction.
Some of those electrons gain greater momentum and move further
in the $+x$ direction.
After many electrons have moved,
the heavy ions are left, and a large $+$charge occurs there.
The electrons that are located at a position away from the laser irradiation
surface, and are still mostly stationary, move in the $-x$ direction and
flow into this area as they are attracted to this charge.
These flowing electrons also form a new electron peak and are accelerated
in the $+x$ direction by the ponderomotive force.
A cycle occurs in which the electrons are pushed away by the laser pulse in the
$+x$ direction and the electrons that are located further away flow into there,
and then next these flowing electrons are pushed away.
The electrons in each region gradually become uniformly mixed,
and the kinetic energy of the electrons of all regions increases.
We call this mechanism `stir-up electron heating' (SEH).
This phenomenon starts early in the laser ion acceleration process, and
the SEH process speed is much faster than the ion acceleration process speed.
From $t=-17$ fs,
the difference in the electron distribution by region disappears,
and all the regions are almost equally mixed.
The peak of the all the electron densities is located near the center of
the initial target in a thickness direction at $t=-17$ fs.
Over time: $t=-17$ to $-7$ fs,
all the electrons and the peak of the electron density move in the $+x$
direction, and the peak of the electron density gradually decreases.

The positional relationship between the ions and electrons is considered.
As shown at $t=-27$ fs in Fig. \ref{fig:fig-ini_e}(a),
the electrons near the surface$^-$ move in the $+x$ direction,
and the ions are left there (the dark blue area).
That is, a layer of ions with a strong $+$charge is formed there.
Looking at the number density distribution of all the electrons at this time,
the black dotted line in Fig. \ref{fig:fig-ini_e}(b), a high peak of the
electron density is present next to the $+x$ side of that ion layer, and
the $+x$ outside area of this peak remain almost its initial density.
That is, a layer of electrons with a strong $-$charge is formed.
Therefore, two layers, a thin ion layer with a strong $+$charge and
a thin electron layer with strong $-$charge, are formed adjacent to each other,
and these are strongly pulled toward each other.
This electron layer is pushed strongly in the $+x$ direction by
the ponderomotive force of the laser pulse.
Therefore, although this high-density electron layer receives
a strong Coulomb force in the $-x$ direction, by the $+$charge ion layer,
it does not move in the $-x$ direction due to the laser pulse
coming from the $-x$ side.
Then, the ion layer near to the surface$^-$ is pulled strongly by
the electron layer which moves fast in the $+x$ direction,
and is strongly accelerated in the $+x$ direction.
That is, the ions of region-1 near surface$^-$ are strongly accelerated in
the $+x$ direction.
At this time, since the density distributions of all the electrons
in the other areas are almost initial density,
such acceleration has not appeared for the ions of the other regions.
Although a peak of electron density appears near the boundary between the
initial regions-2 and 3 at $t=-17$ fs,
and near initial regions-3 and 4 at $t=-7$ fs, those peaks are much smaller
than that of $t=-27$ fs (Fig. \ref{fig:fig-ini_e}(b)).
Therefore, region-3, 4, and 5 ions do not experience
such strong acceleration as region-1 ions.
This is the reason why the ions of region-1 obtain the highest energy and
are distributed on the outermost $+x$ side surface of the target.
That is, the laser pulse forms a very high-density layer of electrons on the
laser irradiation area and if this peak density is highest
during the acceleration process, NH distribution occurs.
Next, the positional relationship between the density distributions of
the ions and electrons and the electric field is shown.

\begin{figure}[tbp]
\includegraphics[clip,width=17.0cm,bb=5 40 523 327]{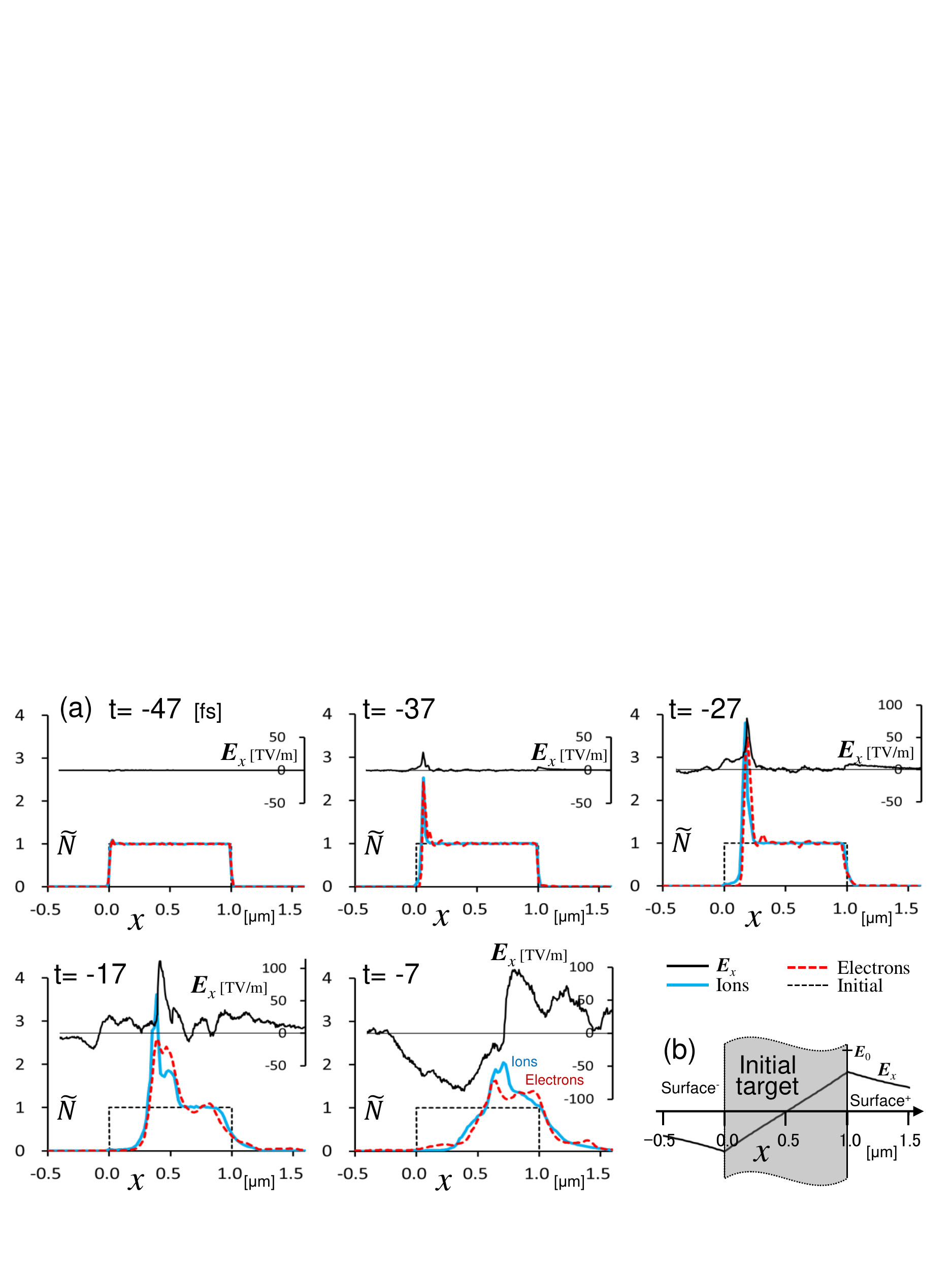}
\caption{
(a) The $x$-component of the electric field, $E_x(x)$,
on the $x$ axis near the target.
Distributions of the number of electrons (red dotted lines),
ions (light blue lines),
and the initial distributions (black dotted lines) which are normalized by
those initial values, $\tilde{N}=N/n_0$.
The peaks of electron and ion number densities move in the $+x$ direction.
The electron number density peaks are slightly ahead of those of the ions.
The $E_x(x)$ peak appears at the peak point of the number density.
Some region-1 ions always position at a high $E_x(x)$ point.
(b) The initial target area in the $x$ direction is $x= 0-1 \mu$m.
$E_x$ is the electric field of the charged cylinder,
and $E_0=\rho\ell/2\epsilon_0$.
}
\label{fig:fig-ini_ei}
\end{figure}

Figure \ref{fig:fig-ini_ei}(a) shows the $x$-component of the electric field,
$E_x(x)$, on the $x$ axis around the target, $-0.5\mu$m $< x < 1.5 \mu$m,
and the number of density distribution of electrons and ions
at the early simulation times $t=-47$ to $-7$ fs, which are the same times
as in Figs. \ref{fig:fig-ini_i} and \ref{fig:fig-ini_e}.
The number of density distribution lines of ions and electrons are the same as
the lines of all the ions and electrons
in Figs. \ref{fig:fig-ini_i}, \ref{fig:fig-ini_e}.
The initial number of density distributions of ions and electrons are indicated
by black dotted lines.
The initial target area is $x= 0-1 \mu$m (see Fig. \ref{fig:fig-ini_ei}(b)).
The density distributions of electrons and ions are approximately the same
at all times, however, the electron density peaks are slightly ahead of those
of the ions in the $+x$ direction in detail,
and the peaks of the electric field are also present there in a similar shape.
These peaks move in the $+x$ direction over time.
At $t=-47$ fs, although the electron and ion distributions remain almost
its initial state, there are a very small peaks near the surface$^-$.
The electric field, $E_x(x)$, is almost zero throughout,
although very small vibrations occur in this part.
At $t=-37$ fs, the distributions of electrons and ions have a sharp peak
near the surface$^-$ (the area slightly inside the target).
And, at the same position, $E_x (x)$ also has a peak of positive values.
That is, a force of acceleration in the $+x$ direction is acting on
this peak of ions.
At $t=-27$ fs, the peaks of the density distribution of electrons and
ions are even higher,
and these peaks move further in the $+x$ direction than the previous time.
And also, $E_x(x)$ has a high peak at that position.
The peak position of the electron density is slightly shifted in the $+x$
direction from that of the ion in $-37 \le t \le -17$ fs.
Therefore, in this time, the bunch of electrons which has a $-$charge attracts
the ion with a $+$charge by the Coulomb force.
These electrons form a high-density bunch and are accelerated in the $+x$
direction by the ponderomotive force, and the ions are
pulled by this bunch of electrons and are accelerated in the $+x$ direction.
Although the accelerating electron bunch is located slightly ahead of the
accelerating ions in the accelerating direction, i.e., $+x$ direction,
the distance is very short.
This is known as radiation pressure acceleration (RPA).
In $t \le -17$ fs,
the electric field has approximately $+$values in the viewed area,
and the strong $+x$ direction electric fields appear
at the density peak position of ions.
That is, the ions in those peaks have been continually and strongly accelerated
in the $+x$ direction during this time.
As time goes by: $t=-17$ to $-7$ fs, the peaks of electrons and ions move
further in the $+x$ direction and those peak densities gradually decrease,
and the sharp peaks become wide.
In the electric field too, the narrow peak of $E_x(x)$ becomes wider and
those are distributed over a wide area.

At $t=-7$ fs,
$E_x(x)$ has a slightly different distribution than the other times.
At the previous times, the electric fields are almost in $+$value across
the display area and, moreover, sharp peaks are present.
On the other hand, at this time, there is a wide high $+$value part and
a wide low $-$value part,
and those have approximately the same wide and absolute values.
$E_x(x)$ has a $+$value on the $+x$ side of the peaks of the
density of ions and electrons,
and has a $-$value on that $-x$ side.
The absolute value of $E_x(x)$ gradually decreases as the distance from
the target surfaces, both the $+$ side and $-$ side, increases.
This electric field distribution, especially on the $-x$ side,
is that of the charged disk and cylinder.
That is, an ion acceleration scheme due to the electric field of
the charged disk has appeared strongly from this time.
Therefore, at this time, a change in the acceleration scheme from the RPA to
the acceleration by the charged disk starts.
The ions are most strongly accelerated at the target surface,
on both $+$ and $-$ sides, in charged disk acceleration, as shown below.
Some ions of region-1 are around $x=1.0 \mu$m, as shown at
$t=-7$ fs in Fig. \ref{fig:fig-ini_i},
and they are moving in the $+x$ direction at about twice the velocity of
the region-5 ions present around them.
Then after this,
these region-1 ions reach the outermost $+x$ side of the target.
Therefore, at the time when the RPA has almost finished, and
the charged disk acceleration starts and becomes the main acceleration scheme,
the region-1 ions are positioned with the velocity, i.e., energy, at the
surface$^+$ which is the most effectively accelerated position thereafter.
Then, these ions are accelerated most effectively by charged disk acceleration,
and obtain more energy, becoming maximum energy ions.
Here, the target consists only of hydrogen;
therefore, the acceleration by the charged disk $=$ the acceleration by
Coulomb explosion of the target.

To summarize the above.
In the initial stage of the acceleration process,
the ions near the laser irradiation surface are strongly accelerated by RPA
in the laser traveling direction, which is the direction toward the inside of
the target.
These accelerated ions pass through the ions and electrons
in the other areas which are almost charge neutral,
and are always located at the position where the high $E_x(x)$ occurs which
moves at high speed in the $+x$ direction,
and keep receiving strong acceleration, and become increasingly high in energy.
Then, those ions appear as high energy on the opposite side surface
of the target, i.e., surface$^+$.
And then, those ions are effectively and further accelerated by the Coulomb
explosion, which becomes the main acceleration scheme around this time.
As a result those ions are at maximum energy.
This is NHA.
The ions located near the surface$^-$ obtain both RPA and Coulomb explosion
effects, most effectively compared to the other ions in the target in NHA.
The obtained energy of those ions is by PRA $+$ Coulomb explosion.
NHA is caused by RPA and subsequent Coulomb explosion of the target.
This is the reason for the occurrence of NHA.

The accelerated ion energy, $\mathcal{E}_i$, is proportional to the laser
energy, $\mathcal{E}_\mathrm{las}$, in RPA \cite{ESI}.
That is proportional to $\sqrt{\mathcal{E}_\mathrm{las}}$ by
Coulomb explosion accelerating, as shown below.
Since NHA $=$ RPA $+$ Coulomb explosion,
$\mathcal{E}_i$(NHA) $=k_\mathrm{R} \mathcal{E}_\mathrm{las} + k_\mathrm{C} \sqrt{\mathcal{E}_\mathrm{las}}$,
where $k_\mathrm{R}$ and $k_\mathrm{C}$ are proportionality constants;
therefore, the accelerated ion energy is proportional to
$\mathcal{E}_\mathrm{las}$, $I$, in NHA.

Next, HA is considered.
This is considered a state in which some electrons are removed from the area
within the laser spot diameter of the foil target by laser irradiation.
At this time, the foil target is assumed as a charged disk or cylinder with
a diameter equal to the laser spot size.
The electric field of the charged cylinder is considered.
It is assumed that the $x$ axis is in the cylinder height direction,
and the origin is at the center of the cylinder whose
radius is $R$ and height is $\ell$.
The $x$-component of the electric field on the $x$ axis
in the charged cylinder is written as \cite{TM3}
\begin{equation}
E(x)= \frac{\rho\ell}{2\epsilon_0} \bigl[a(x)+b(x) \bigr],
\label{exab}
\end{equation}
where $\rho$ is the charge density, $\epsilon_0$ is the vacuum permittivity,
and
\begin{equation}
a(x)=
\sqrt{\Bigl(\frac{x}{\ell}-\frac{1}{2}\Bigr)^2+\Bigl(\frac{R}{\ell}\Bigr)^2}
-\sqrt{\Bigl(\frac{x}{\ell}+\frac{1}{2}\Bigr)^2+\Bigl(\frac{R}{\ell}\Bigr)^2},
\label{exa}
\end{equation}
$b(x)$ in the cylinder, $-\ell/2 \leq x \leq \ell/2$, is
\begin{equation}
b(x)=
\frac{2x}{\ell}.
\label{exb}
\end{equation}
The function $E(x)$ passes through the origin, $E(0)=0$, and a monotonically
increasing function in $-\ell/2 \leq x \leq \ell/2$
(see Fig. \ref{fig:fig-ini_ei}(b)).
Therefore, in this charged target, $E(x)$ has $+$value for the $+x$ side region
from the center and has $-$value for the $-x$ side region,
and the absolute values increase as it goes to both surfaces from the center.
In this electric field,
ions in the target are accelerated more strongly toward the outside of the
target as their position go from the center to the surfaces of the target,
which are both surface$^-$ and surface$^+$.
That is, as the initial position of the ion locates away from the target
center in the $+x$ direction, the ion is strongly accelerated in the $+x$
direction, and also strongly accelerated in the $-x$ direction as it locates
away in the $-x$ direction.
The shape of this electric field does not change even if the target expands.
That is,
in this situation, as the position of the ion is closer to the surface of the
target, the ion is strongly accelerated during the Coulomb explosion process.
This is known as a Coulomb explosion acceleration scheme.
This situation occurs when the laser intensity, energy, is relatively weak,
or the number of electrons in the thickness direction is great relative to
the laser intensity, energy.
This is HA.

The relation between laser energy, $\mathcal{E}_\mathrm{las}$, and
obtained ion energy, $\mathcal{E}_{i}$, in the disk target
is written as \cite{MBEKK}
\begin{equation}
  \mathcal{E}_\mathrm{las}=
  \frac{\pi \epsilon_0 R\mathcal{E}_{i}^2}{q_{i}^2},
\label{elas}
\end{equation}
where $q_{i}$ is the ion charge.
A model is used in which an ion is placed on the opposite side surface of the
laser irradiation surface and the ion is accelerated from the stationary state
by the electric field of the charged disk.
Hence, it is an HA situation.
We solve the equation for $\mathcal{E}_{i}$, then we get
$\mathcal{E}_{i} = k \sqrt{\mathcal{E}_\mathrm{las}}$,
where $k = q_{i} / \sqrt{\pi\epsilon_0 R}$ and
it is the same value in all cases in our simulations.
Furthermore, according to Eq. (\ref{els}),
$\mathcal{E}_\mathrm{las}=I \cdot V^*$ and $V^*$ is a constant value.
Therefore, in HA, the obtained ion energy is proportional to
$\sqrt{I}$, $\sqrt{\mathcal{E}_\mathrm{las}}$.
The surface charge density, $\rho\ell$, is important for obtained ion energy
in HA, as shown in Eq. (\ref{exab}).

Whether the target is thick or thin, if the number of electrons taken out
per target surface unit area are the same,
then $\rho\ell$ will be the same and there will be no difference
in the obtained ion energy.
Therefore, the difference in target thickness does not significantly affect
the obtained ion energy in HA.

NHA is due to the acceleration scheme being RPA $+$ Coulomb explosion, and
HA is due to the mainly Coulomb explosion only being dominant in that.

\section{-CH$_2$- foil target} \label{ch2}

In the previous section, it is shown that NHA appears
when the high energy laser irradiates on the hydrogen foil target.
Here, we show that NHA does not occur only in the hydrogen target.
The simulation results using a polyethylene (-CH$_2$-) foil target are shown.
The polyethylene (-CH$_2$-) target is chosen because -CH$_2$- is easy to handle
since the material is solid at room temperature,
and can produce high energy protons as well as the hydrogen target \cite{TM2}.

The parameters used in simulation of the polyethylene (-CH$_2$-) foil
target are given.
Since many parameters are the same as the hydrogen foil target case
(Sec. \ref{h-resu}),
the different points of their conditions are shown below.
The laser intensity and energy chosen are $I = 1 \times 10^{22}$ W/cm$^2$ and
$\mathcal{E}_\mathrm{las}=25$ J, respectively,
which are the conditions shown the results in detail in the previous section.
The other laser conditions are the same as in the previous section.
The foil thickness is set to be a thickness that gives the same areal electron
number density, $\int n_{e} dx$, as $1.0 \mu$m hydrogen foil.
Here, the foil thickness is simply set to $0.1 \mu$m,
although it is strictly $0.16 \mu$m.
The ionization state of a carbon ion is assumed to be $Z_{i}=+6$.
The electron density is $n_{e}=3\times 10^{23}$ cm$^{-3}$.
Therefore, the proton density is $n_{e}/4$
and the carbon ion density is $n_{e}/8$.
The total number of quasiparticles is $1\times 10^{9}$.
The number of grid cells is equal to $55000\times55000$
along the $X$, and $Y$ axes, respectively.
Correspondingly, the simulation box size is $102 \mu$m $\times 102 \mu$m.
The laser-irradiated side surface of the foil is placed at $X=40 \mu$m,
and the center of the laser pulse is located $20 \mu$m behind it.

\begin{figure}[tbp]
\includegraphics[clip,width=8.0cm,bb=1 1 430 479]{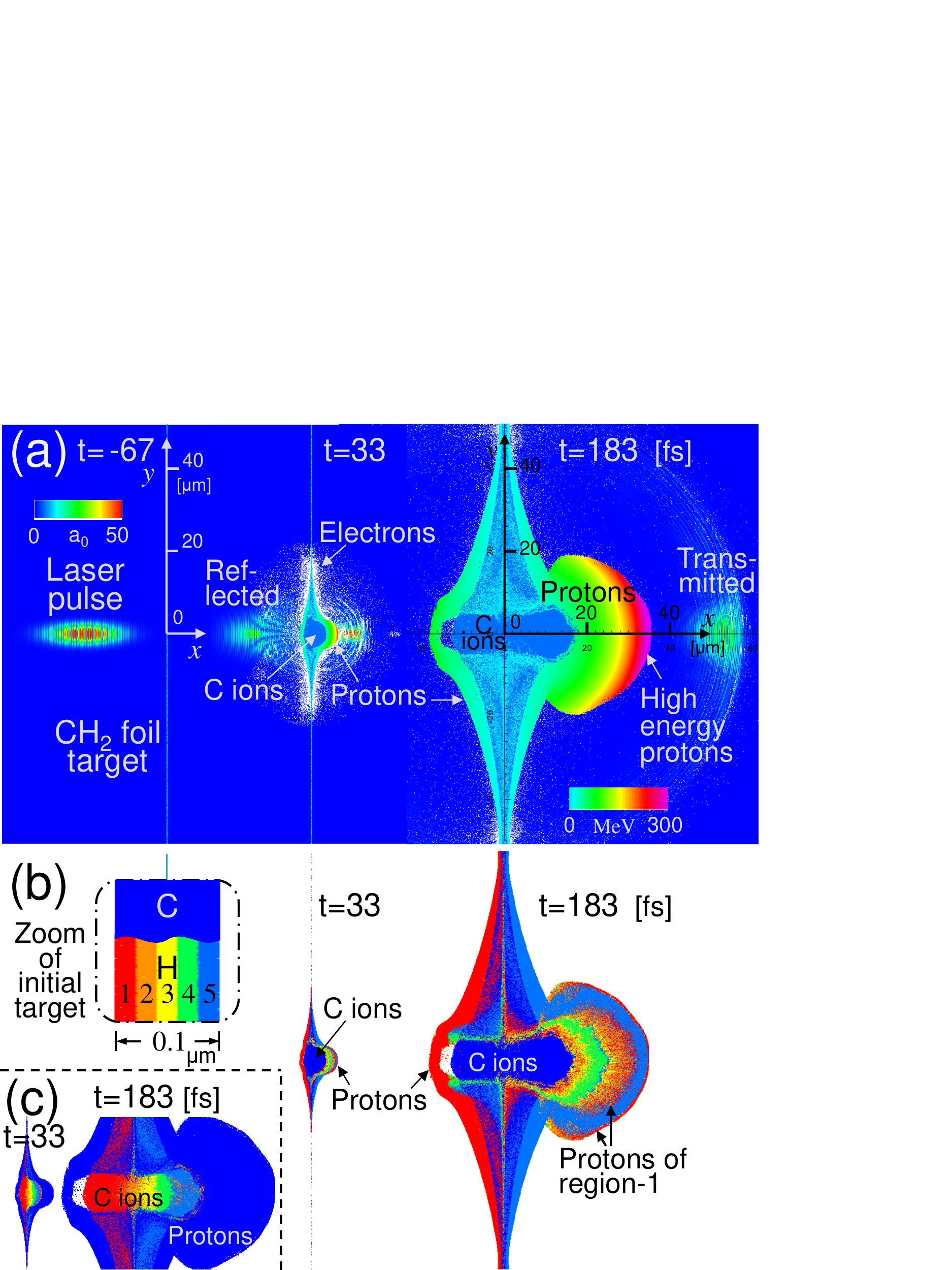}
\caption{
Spatial distribution of particles and electric field magnitude of each time
in the -CH$_2$- foil target case
which is $I=1 \times 10^{22}$ W/cm$^2$ ($\mathcal{E}_\mathrm{las}=25$ J) and
$\ell=0.1\mu$m.
(a) For protons, the color corresponds to their energy,
(b) the color corresponds to their initial positon.
The protons on outermost part of the $+x$ side and spread around the area on
the $+x$ side of the accelerated proton cloud come from near surface$^-$.
The NHA regime also appears in  the -CH$_2$- target.
(c) For carbon ions, the color corresponds to their initial position.
A few carbon ions from near to surface$^-$ appear in
the outermost $+x$ direction part of the carbon ion cloud.
}
\label{fig:fig-ch}
\end{figure}

Figure \ref{fig:fig-ch}(a) shows the result of the -CH$_2$- foil target case at each time.
The protons are color-coded by their energy.
The initial state is shown at $t=-67$ fs.
At $t=33$ fs, part of the laser pulse is reflected by the target while
another part is transmitted the target.
-CH$_2$- near the laser irradiation area is separated into a carbon ions
region and a protons region.
The carbon ion cloud distributes near the center of the expanded target,
and the proton cloud distributes around it, so as to cover it.
The target is greatly expanded at $t=183$ fs, and the proton cloud in particular
is largely expanded in the laser propagation direction.
This distribution of protons is similar to the results
in the case of hydrogen foil.
The maximum proton energy at $t=183$ fs is $332$ MeV,
which is almost the same as the hydrogen foil case,
and those appear on the outermost part of the $+x$ side of the expanded target.
Proton energy gradually decreases as goes in the $-x$ direction from
there in the range of $x>0$.
The maximum carbon ion energy is $54$ MeV/u,
which is about 1/6 of the maximum proton energy.
Although Fig. \ref{fig:fig-ch}(b) is the same result as
Fig. \ref{fig:fig-ch}(a), the protons are color-coded according
to their initial positions.
The initial foil is sliced evenly into five in the thickness direction,
i.e., $x$ direction, and each region of protons has a different color,
although the carbons are not color-coded
(see zoomed initial target in Fig. \ref{fig:fig-ch}(b)).
At $t=183$ fs, it shows that the protons near by the $+x$ side outermost part
come from near the opposite, i.e., the $-x$, side surface.
That is, the high energy protons are the protons that are initially located
near the laser-irradiated surface, surface$^-$.
That is the NHA regime too.
Also, the protons near the surface$^-$ have already appeared on the outermost
part of the $+x$ side of the target, at $t=33$ fs, indicating that
this distribution is formed at the early stage of the acceleration process.
These are the same as in the hydrogen foil case.
Fig. \ref{fig:fig-ch}(c) is the same result as Fig. \ref{fig:fig-ch}(a)(b),
although the carbon ions are color-coded according to their initial positions.
Although the carbon ions are almost HA,
we can see that some carbon ions from region-1 appear at the outermost
area of the carbon ion cloud in the $+x$ direction.

NHA is not a special phenomenon that occurs only in hydrogen foil targets,
but also occurs in polyethylene (-CH$_2$-) foil.

\section{Conclusions}

Particle acceleration driven by a laser pulse irradiating a thin-foil target is
investigated with the help of 2D PIC simulations.
We considered the laser particle acceleration from a topology viewpoint.
The acceleration schemes are classified into homeomorphic acceleration (HA) and
non-homeomorphic acceleration (NHA), according to whether the spatial
distribution of accelerated ions is homeomorphic
to the initial target or not.
Ion acceleration by RPA $+$ Coulomb explosion is classified into NHA, and
the acceleration by mainly only Coulomb explosion is classified into HA, under
the conditions of our simulations.
In the NHA regime,
the obtained proton energy rises in proportion to the laser intensity $I$,
laser energy $\mathcal{E}_\mathrm{las}$,
and the foil thickness strongly affects the obtained ion energy.
On the other hand, in the HA regime,
the obtained proton energy rises in proportion to
$\sqrt{I}$, $\sqrt{\mathcal{E}_\mathrm{las}}$,
and the foil thickness does not affect this much.
The HA regime is not efficient in producing high energy ions.
In order to generate the highest energy ions as possible at a given laser
capacity, i.e., generate high energy ions efficiently,
it is necessary to produce the NHA regime.
Moreover, it is important to use the optimal target thickness in this regime.
We have found that the laser acceleration schemes are classified
and elucidated using topology.

\section*{Acknowledgments}
The computations were performed using the ICE X supercomputer at JAEA Tokai.


\begin{thebibliography}{99}

\bibitem{BWE}
S. V. Bulanov, J. J. Wilkens, T. Esirkepov, G. Korn, G. Kraft, S. D. Kraft,
M. Molls, and V. S. Khoroshkov: Phys. - Usp. \textbf{57}, 1149 (2014)

\bibitem{DNP}
H. Daido, M. Nishiuchi, and A. Pirozhkov:
Rep. Prog. Phys. \textbf{75}, 056401 (2012)

\bibitem{CLK}
E. L. Clark, K. Krushelnick, J. R. Davies, M. Zepf, M. Tatarakis, F. N. Beg,
A. Machacek, P. A. Norreys, M. I. K. Santala, I. Watts and A. E. Dangor:
Phys. Rev. Lett. \textbf{84}, 670 (2000).

\bibitem{SNV}
R. A. Snavely, M. H. Key, S. P. Hatchett, T. E. Cowan, M. Roth, T. W. Phillips,
M. A. Stoyer, E. A. Henry, T. C. Sangster, M. S. Singh, S. C. Wilks,
A. MacKinnon, A. Offenberger, D. M. Pennington, K. Yasuike, A. B. Langdon,
B. F. Lasinski, J. Johnson, M. D. Perry and E. M. Campbell:
Phys. Rev. Lett. \textbf{85}, 2945 (2000).

\bibitem{ROT}
M. Roth, T. E. Cowan, M. H. Key, S. P. Hatchett, C. Brown, W. Fountain,
J. Johnson, D. M. Pennington, R. A. Snavely, S. C. Wilks, K. Yasuike, H. Ruhl,
F. Pegoraro, S. V. Bulanov, E. M. Campbell, M. D. Perry, and H. Powell,
Phys. Rev. Lett. \textbf{86}, 436 (2001).

\bibitem{ESI}
T. Esirkepov, M. Borghesi, S. V. Bulanov, G. Mourou, and T. Tajima,
Phys. Rev. Lett. \textbf{92}, 175003 (2004).

\bibitem{BEE}
S. V. Bulanov, E. Yu. Echkina, T. Zh. Esirkepov, I. N. Inovenkov, M. Kando,
F. Pegoraro, and G. Korn,
Phys. Rev. Lett. \textbf{104}, 135003 (2010).

\bibitem{BWP}
J. Badziak, E. Woryna, P. Parys, K. Yu. Platonov, S. Jablo\'{n}ski,
L. Ry\'{c}, A. B. Vankov, and J. Woowski,
Phys. Rev. Lett. \textbf{87}, 215001 (2001).

\bibitem{DL}
T. Esirkepov, S. V. Bulanov, K. Nishihara, T. Tajima, F. Pegoraro,
V. S. Khoroshkov, K. Mima, H. Daido, Y. Kato, Y. Kitagawa, K. Nagai,
and S. Sakabe,
Phys. Rev. Lett. \textbf{89}, 175003 (2002).

\bibitem{HSM}
M. Hohenberger, D. R. Symes, K. W. Madison, A. Sumeruk, G. Dyer, A. Edens,
W. Grigsby, G. Hays, M. Teichmann, and T. Ditmire,
Phys. Rev. Lett. \textbf{95} 195003 (2005).

\bibitem{PPM}
F. Peano, F. Peinetti, R. Mulas, G. Coppa, and L. O. Silva,
Phys. Rev. Lett. \textbf{96}, 175002 (2006).

\bibitem{PRK}
A. P. L Robinson, A. R. Bell, and R. J. Kingham,
Phys. Rev. Lett. \textbf{96}, 035005 (2006).

\bibitem{SPJ}
H. Schwoerer, S. Pfotenhauer, O. J\"{a}ckel, K.-U. Amthor, B. Liesfeld,
W. Ziegler, R. Sauerbrey, K. W. D. Ledingham, and T. Esirkepov,
Nature (London) \textbf{439}, 445 (2006).

\bibitem{Toncian}
T. Toncian, M. Borghesi, J. Fuchs, E. d'Humi\`{e}res, P. Antici, P. Audebert,
E. Brambrink, C. A. Cecchetti, A. Pipahl, L. Romagnani, and O. Willi,
Science {\bf 312}, 410 (2006).

\bibitem{TM1}
T. Morita,
Phys. Plasmas \textbf{20}, 093107 (2013).

\bibitem{TM2}
T. Morita,
Phys. Plasmas \textbf{21}, 053104 (2014).

\bibitem{CBL}
C. K. Birdsall and A. B. Langdon,
\textit{Plasma Physics via Computer Simulation} (McGraw-Hill, New York, 1985).

\bibitem{Kiri}
H. Kiriyama et al., Opt. Lett. {\bf 43}, 2595 (2018).

\bibitem{TM3}
T. Morita,
Phys. Plasmas \textbf{24}, 083104 (2017).

\bibitem{MBEKK}
T. Morita, S. V. Bulanov, T. Zh. Esirkepov, J. Koga, and M. Kando,
J. Phys. Soc. Jpn. \textbf{81}, 024501 (2012).

\end{thebibliography}
\end{document}